 \documentclass[final,5p,times]{elsarticle}

\usepackage{lineno}
\usepackage{url}
\usepackage{graphicx}
\usepackage{amstext,amsmath,amssymb,amsthm}
\usepackage{upgreek}
\usepackage{enumerate}
\usepackage{txfonts}
\usepackage{graphicx}
\usepackage{subfigure}
\usepackage{multirow}
\usepackage{multicol}
\usepackage[T1]{fontenc}

\begin{document}

\begin{frontmatter}



\title{Ground-based calibration and characterization of the HE detectors for \emph{Insight}-HXMT}


\author[IHEP]{XuFang Li}
\author[IHEP]{CongZhan Liu\corref{cor1}}\ead{liucz@ihep.ac.cn}
\author[IHEP]{Zhi Chang}
\author[IHEP]{YiFei Zhang}
\author[IHEP]{XiaoBo Li}
\author[IHEP,UCAS]{He Gao}
\author[IHEP]{ZhengWei Li}
\author[IHEP]{XueFeng Lu}
\author[IHEP]{Xu Zhou}
\author[IHEP]{Aimei Zhang}
\author[IHEP]{Tong Zhang}
\author[IHEP]{FangJun Lu}
\author[IHEP]{YuPeng Xu}
\author[IHEP,UCAS]{ShuangNan Zhang}
\author[IHEP,UCAS,Tsinghua]{TiPei Li}
\author[IHEP]{Mei Wu}
\author[IHEP]{Shu Zhang}
\author[IHEP]{HongWei Liu}
\author[IHEP]{Fan Zhang}
\author[IHEP]{LiMing Song}
\author[Tsinghua]{YongJie Jin}
\author[Tsinghua]{HuiMing Yu}
\author[Tsinghua]{Zhao Zhang}
\author[Tsinghua]{MinXue Fu}
\author[Tsinghua]{YiBao Chen}
\author[Tsinghua]{JingKang Deng}
\author[Tsinghua]{RenCheng Shang}
\author[Tsinghua]{Guoqing Liu}
\author[NIM]{JinJie Wu}
\author[NIM]{HaoRan Liu}
\author[NIM]{JunCheng Liang}
\author[NIM]{XiangPing Qiu}
\cortext[cor1]{Corresponding author. }
\address[IHEP]{Key Laboratory of Particle Astrophysics, Institute of High Energy Physics, Chinese Academy of Sciences, Beijing, China}
\address[UCAS]{University of Chinese Academy of Sciences, Beijing, China}
\address[Tsinghua]{Tsinghua University, Beijing, China}
\address[NIM]{Division of Ionizing Radiation, National Institute of Metrology, Beijing, China}
\begin{abstract}
High energy X-ray telescope (HE) is one of the three instruments of Insight-HXMT(
Hard X-ray Modulation Telescope) payload. The HE detector (HED) array
is composed of 18 actively NaI(Tl)/CsI(Na) phoswich scintillators with a total
geometric area of $\sim5100\,\mathrm{cm^2}$ and cover the energy range 20$\sim$250\,keV. In this paper
we describe the on-ground detector-level calibration campaigns and present the principal
instrument properties of HEDs.

\end{abstract}


\end{frontmatter}


\section{Introduction}
\label{sec:1 introduction}
The high energy X-ray telescope (HE) is one of the three main instruments onboard the Hard X-ray Modulation Telescope (HXMT)(\citep{LiTipei..HXMT..2007},\citep{LiTipei..2013},\citep{LuFangjun..HXMT..2018},\citep{S.Zhang..2018}), which had been launched on June 15th, 2017.

The over HE instrument comprises six components: the high energy X-ray detector (HED) array, their respective collimators and automatic gain control detectors (AGC), anticoincidence shield system, particle monitors and electronic unit. The main design requirements of HE is listed in Table \ref{tab:tab1 features}.

The HED array is composed of 18 NaI(Tl)/CsI(Na) phoswich scintillators with a total geometric area of $5100\,\mathrm{cm^2}$ and with energy band of 20 to 250\,keV.  It is designed to observe spectra, temporal variability of X-ray compact objects in the sky. And also it has capacity of imaging by scanning a specific region based on the Direct Demodulation Method (DDM)(\citep{LiTipei..DDM..1993}). The determination of spectral and temporal properties from HED data requires very detailed knowledge of the full response of HED. And the physical detector response of the HED instrument to $X/\gamma$ rays is determined with the help of Monte Carlo simulations \citep{Geant4..2003}, which are supported and verified by on-ground experimental calibration measurements.

In order to perform the above validations, several pre-flight calibration experiments were carried out at different levels. This paper focuses on the detector-level calibration campaigns of the HED/NaI, and in particular on the analysis methods and results, which crucially support the development of a consistent HED instrument response.

\begin{table}
  \centering
  \caption{Main characteristics of HE}\label{tab:tab1 features}
    \begin{tabular}{ll}
    \hline
    Parameter & Value \\
    \hline
      Energy band & 20-250\,keV  \\
      Total geometric area  & $5100\,\mathrm{cm^2}$  \\
      Main detector    &  NaI(Tl)/CsI(CsI) \\
                        &  $\sim$3.5\,mm/40\,mm \\
      Timing accuracy & $<\,10\,{\mu}s$\\
      Deadtime & $<\,10\,{\mu}s $ \\
      Combined FOV\,(FWHM)  & $5.7^{\circ}{\times}5.7^{\circ}$  \\
      Energy resolution at 60keV    & $\leq 19\%$ \\
      Maximum count rate &  >30,000\,cnts/sec \\
    \hline
    \end{tabular}
\end{table}

\section{Instrument description}
\label{sec:2 instrument description}
 Totally, we built 24 HEDs, including eighteen flight modules(FM) and six spare modules.  First, all the detectors were calibrated with a tunable X-ray monochromator at the China National Institute of Metrology (NIM). According to the first-step calibration results, four modules were rejected because of their low performance. The subsequent calibration campaign with radioactive sources just performed on the twenty detectors with good performance. Note that the detector numbering scheme used in the calibration and adopted throughout this paper is different to the one used for in-flight analysis. And we just discuss the performance of the eighteen flight modules in this paper.

\begin{figure}
	\centering
\includegraphics[width=0.48\textwidth]{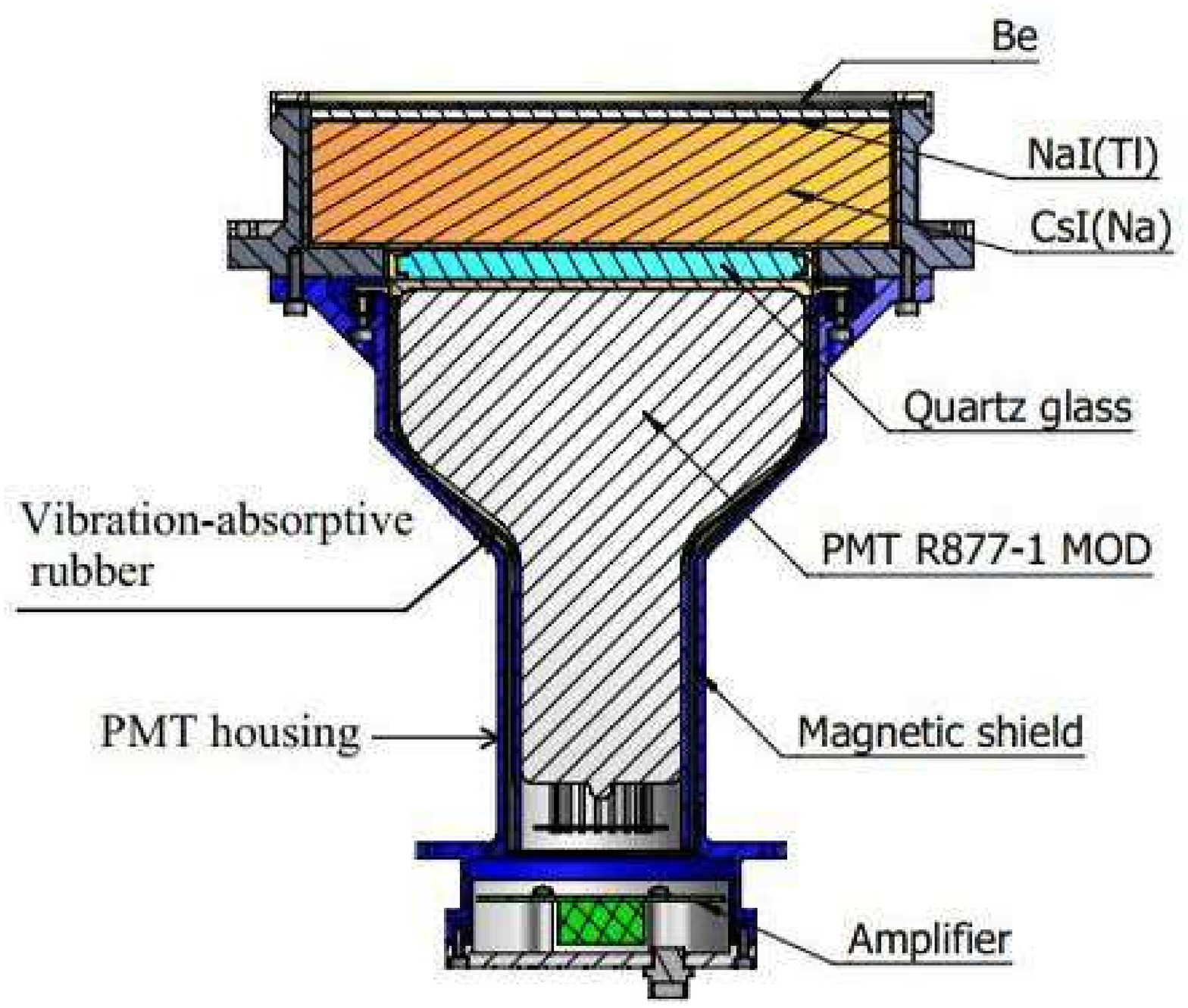}
	\caption{Schematic cross-section of a HED showing the main components}
	\label{FIG:1}
\end{figure}

Each HED (see Figure \ref{FIG:1}) phoswich employs a circular NaI(Tl) primary crystal which is 3.5\,mm thick and 190\,mm diameter. The crystal is optically coupled to a 40\,mm thick and 190\,mm diameter CsI(Na) crystal which is used for active shielding and gamma-ray burst monitor. The coupled crystals are hermetically sealed in an Al-alloy housing with a 1.5\,mm thick Be entrance window and a 10\,mm thick quartz window in the rear. A more ruggedized Hamamatsu R877-01 modified photomultiplier tube is optically coupled to the outside of the quartz window. There is a vibration-absorptive rubber and a permalloy magnetic shield housing between the photomultiplier and the Al housing, then the HED mechanical and magnetic shielding requirements can be fulfilled. The output signals of the PMT are first amplified via the preamplifier of the HED front end electronics. Pulse shape discrimination(PSD) is used to distinguish whether energy absorption occurred in the NaI(Tl) or the CsI(Na) or both \citep{Frontera..PDS..1997}. In the data analysis only the events in the NaI(Tl) were selected. In flight with an active temperature controller, the HED works at $18\pm2{}^{\circ}\mathrm{C}$.


The HE electronic subsystem consists of two boxes (HEA and HEB), which control the instrument and processed data. The HEA is in charge of powering the HEB and the detectors, performing remote control commands, and generating telemetry signals and data. The HEB is in charge of data acquisition (DAQ) and processes signals from all of HE detectors. For each detected event qualified as a good one, the pulse height, pulse shape and the event time would be digitized and be formatted to transmit. In addition to the above information, other housekeeping information is provided by the HEB and transmitted. This includes count rates of every HED, dead time counters, status information, monitoring parameters and so on. The pulses height and pulse shape are digitized into 256 linear channels. Due to the unavailability of the flight HEA and HEB, the engineering qualification models (QM) were used.

In flight the gain of the HED is continuously monitored and automatically controlled by using an automatic gain control detector (AGC). The AGC consists of an $^{\mathrm 241}$Am radioactive source distributed within a plastic scintillator (BC448M) and a multi-pixel photon counter (MPPC, a kind of SiPM product from Hamamatsu, S10362-33-050C). Part of the 59.5\,keV X-rays emitted by the source impinges on the HED. These events are tagged thanks to the 5.5\,MeV $\alpha$ particles simultaneously emitted by the $^{\mathrm 241}$Am radioactive source and detected by the plastic scintillator. In this way, the gain changes are compensated by the HEB software by adjusting the PMT HV to keep the 59.5\,keV line at the specified energy channel 50. During the calibration, an QM AGC was used and the relative location between the AGC and each HED was the same as on board.

\section{Calibration campaign} \label{sec:3 calibration campaign}
To derive the source flux of the observed sources and the source spectrum, a detailed knowledge of the HED response is necessary. Because the response matrix of each detector must be generated using computer simulations, the actual detector response at discrete energies in normal direction of the detector plane has to be measured to verify the validity of the simulated responses. The complete response matrix of the whole HE instrument system is finally created by simulation of a dense grid of energies and infalling photon direction using the verified simulation tool.

The following subsections are dedicated to the description of the calibration campaigns at HED. The most complete calibration of all HED models was performed at a hard X-ray Calibration Facility (HXCF) \citep{ZhouXu..2014}, which was developed for the calibration of HED, with the support and collaboration of the  China National Institute of Metrology (NIM), and allowed the scanning of the detector geometric area with a pencil beam of desired energy over the energy band from 20 to 150\,keV. As a supplement to the HXCF, we also used a set of radioactive sources to perform another calibration campaign to all the detectors in NIM laboratory.

\subsection{Calibration at HXCF} \label{sec:3.1 cal at HXCF}
\begin{figure*}[htb]
  \centering
  \subfigure[]{
  \includegraphics[width=0.45\textwidth]{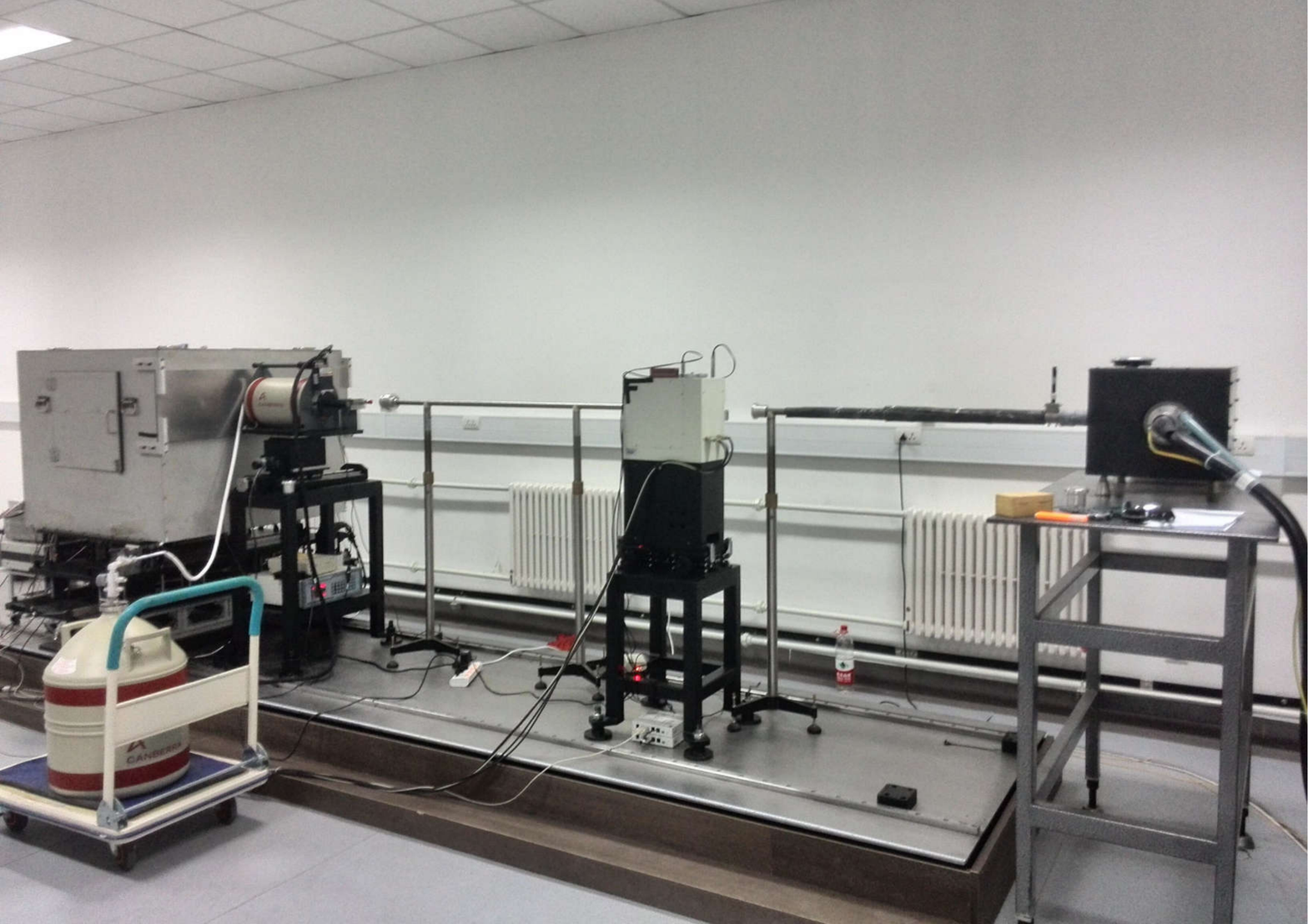}}
  \subfigure[]{
  \includegraphics[width=0.45\textwidth]{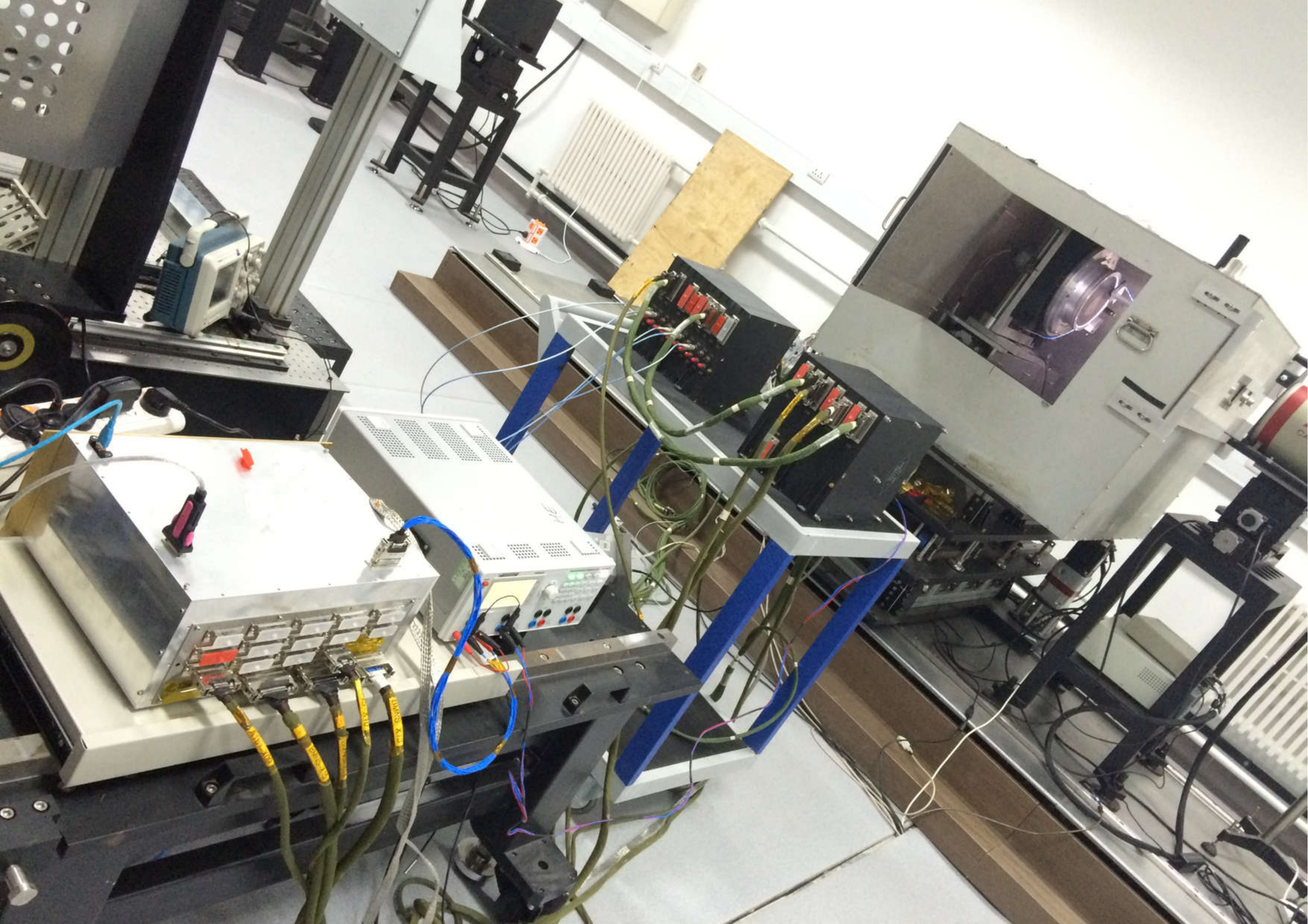}}
  \subfigure[]{
  \includegraphics[width=0.45\textwidth]{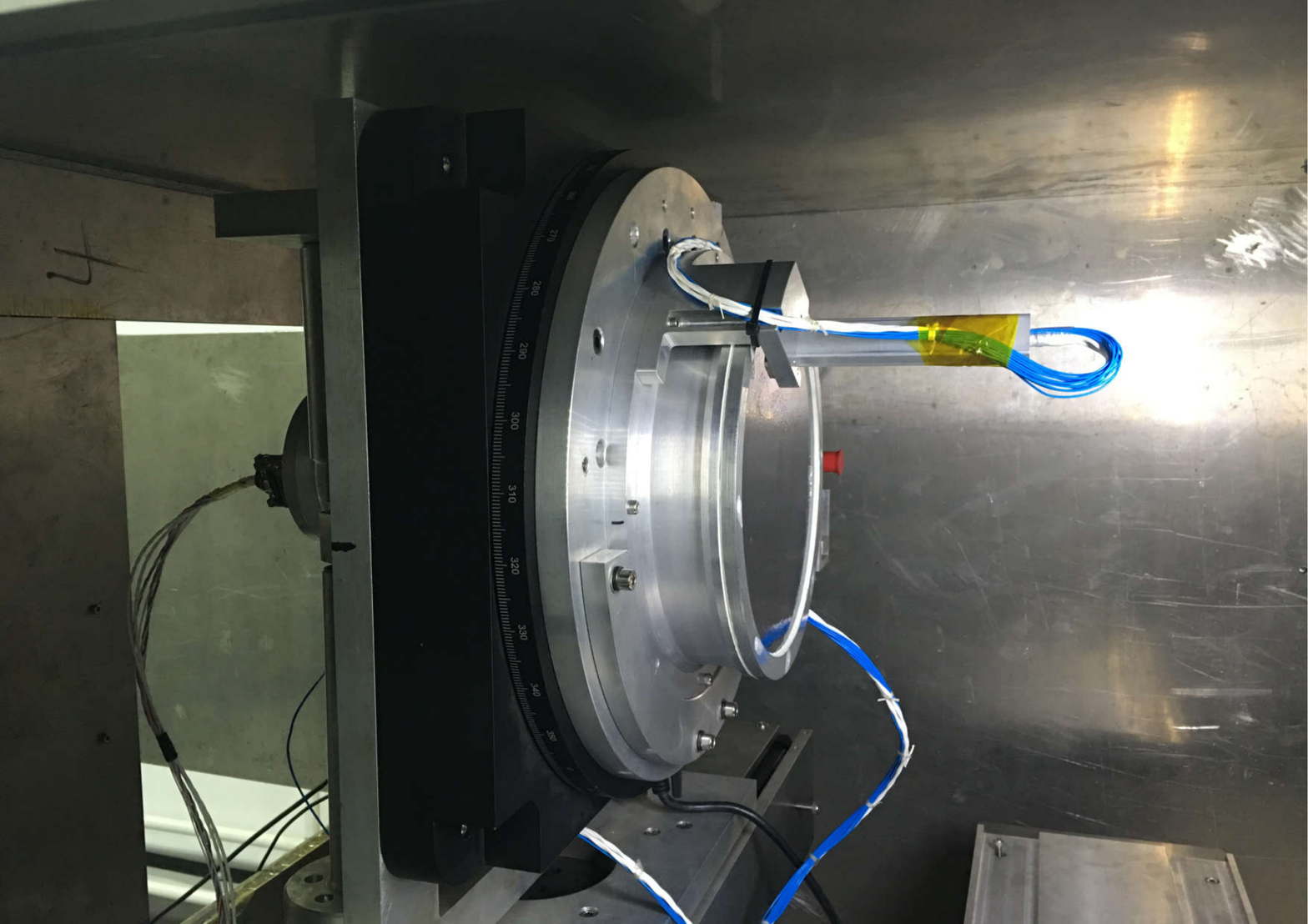}}
  \subfigure[]{
  \includegraphics[width=0.45\textwidth]{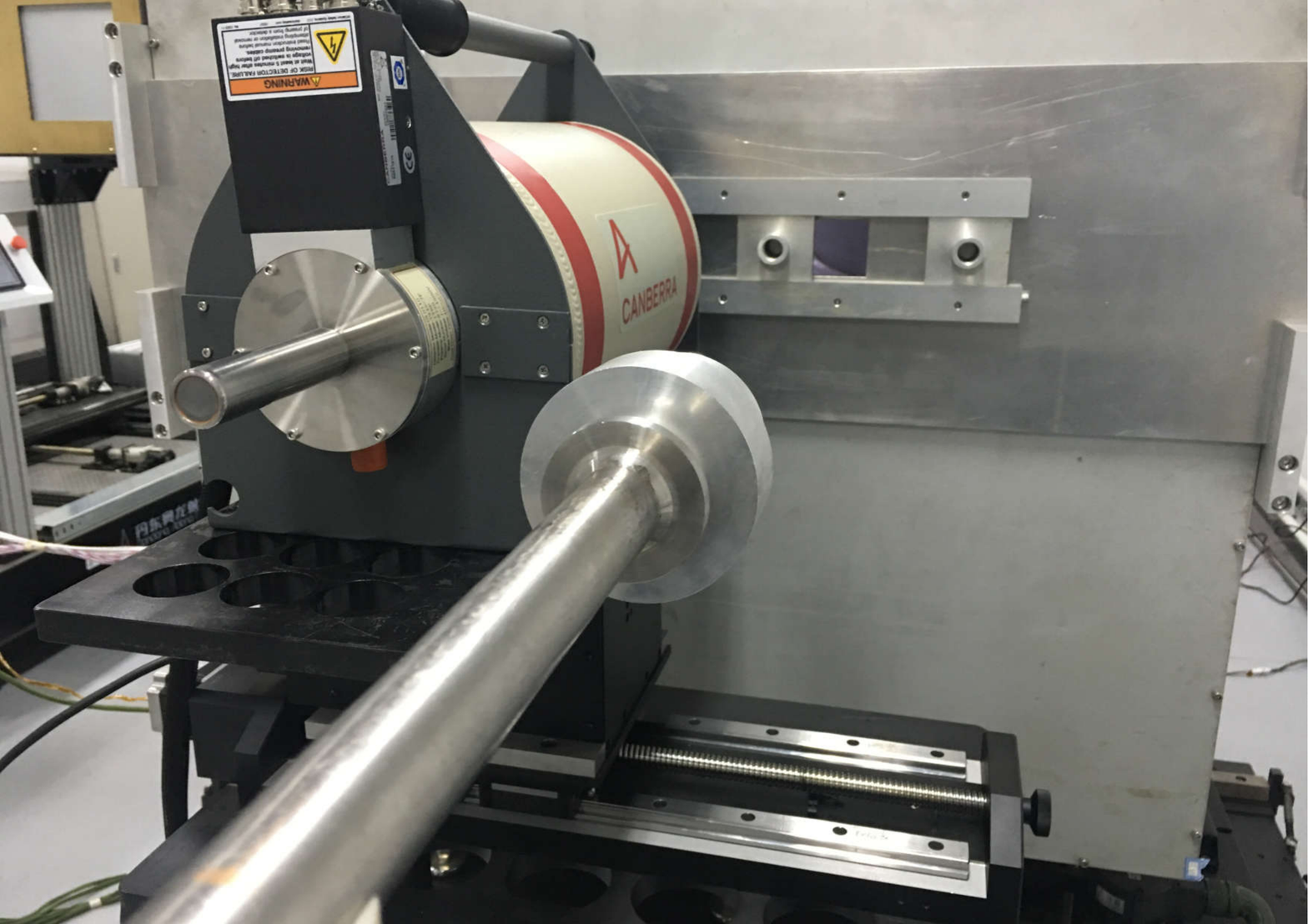}}
  \caption{Pictures of instruments during the calibration campaign at HXCF. (a) X-ray beamline experimental set-up, (b) All of electronics testing system, (c) A HED is mounted on the ($X-\theta$) calibration stand, (d) The LEGe detector is located before the HED shielding box}
  \label{fig:fig2}
\end{figure*}

A description of the HXCF in the first version was reported \citep{ZhouXu..2014}, while after upgrade the energy range has been extended to 15-168\,keV. Some pictures of the calibration setup are shown in figure \ref{fig:fig2}.The main components include an X-ray tube , a double crystal monochromator , a system of collimators , a two-axis (X-Z) table used as standard detector holder , a big shielding box where a two-axis ($X-\theta$) stage as the HED holder is installed. The box is aluminum-lead-aluminum sandwich structure and can stop the scattered radiation in the testing room and reduce the background. The environmental temperature of the lab was set at $18\pm2{}^{\circ}\mathrm{C}$.

The collimator aperture before the first crystal was $\O$2\,mm in diameter. From the crystal spread and divergence of the polychromatic X-ray beam impinging on the crystal 1, the expected energy width   of the exit beam ranges from 0.1 to 6.9\,keV (FWHM) in the 20-150\,keV band, respectively. After the monochromator there was a similar collimator positioned and the final aperture close to the detector was $\O$4\,mm. Considering the limit of data processing capacity of the DAQ box, the flux of the monochromatic light during calibration campaigns were controlled to less than five thousand photons per second.

A low energy germanium detector (LEGe) as the standard detector was calibrated with a set of radioactive sources and its quantum detection efficiency (QDE) of unity had been determined earlier \citep{Liuhaoran..2016}. So with the standard LEGe detector, we could characterize the monochromatic X-ray beam, such as intensity, intrinsic energy width and average energy, at each specific energy point.

\begin{figure*}[!htbp]
  \centering
  \subfigure[]{
  \includegraphics[width=0.3\textwidth]{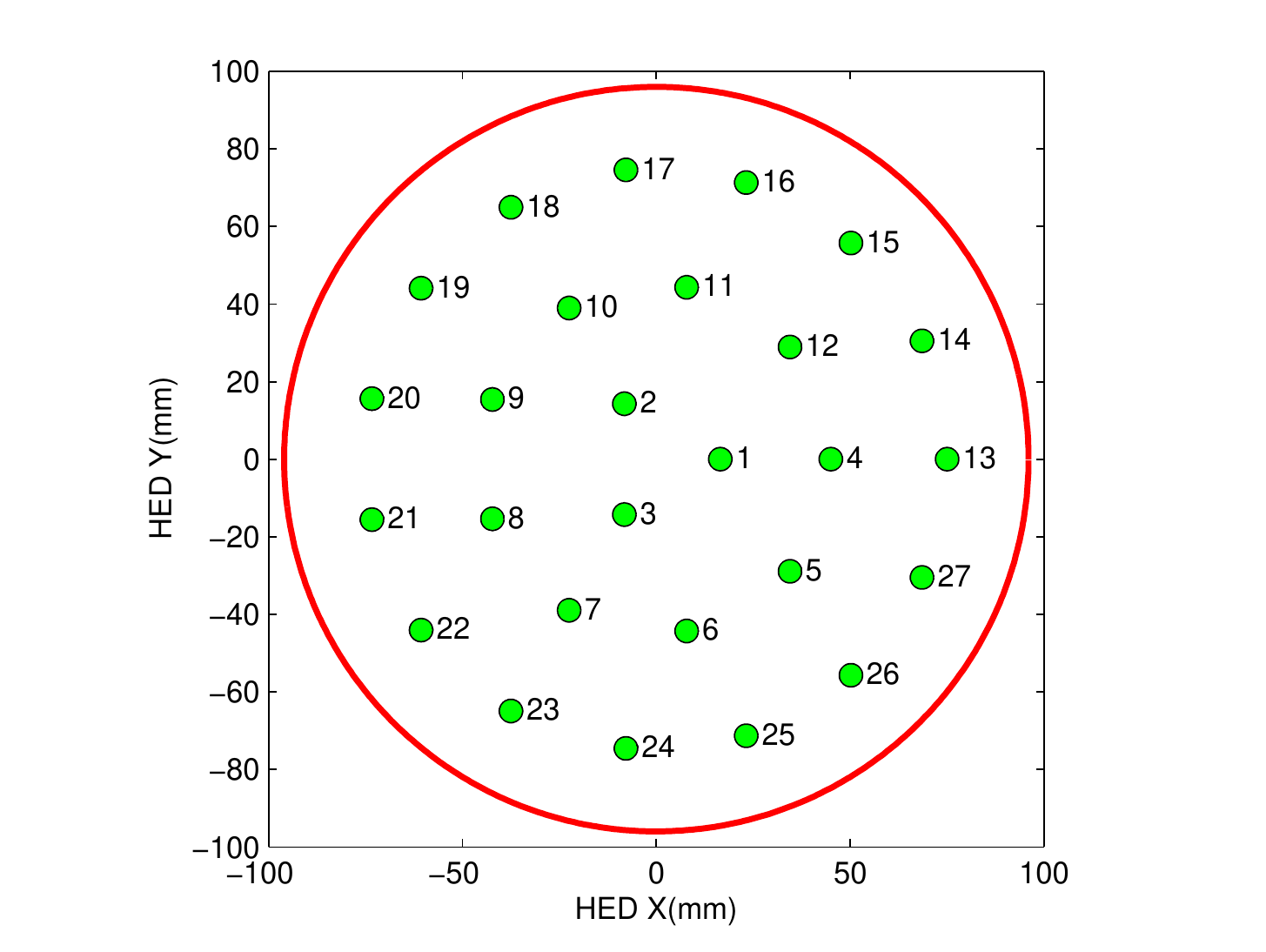}}
  \subfigure[]{
  \includegraphics[width=0.3\textwidth]{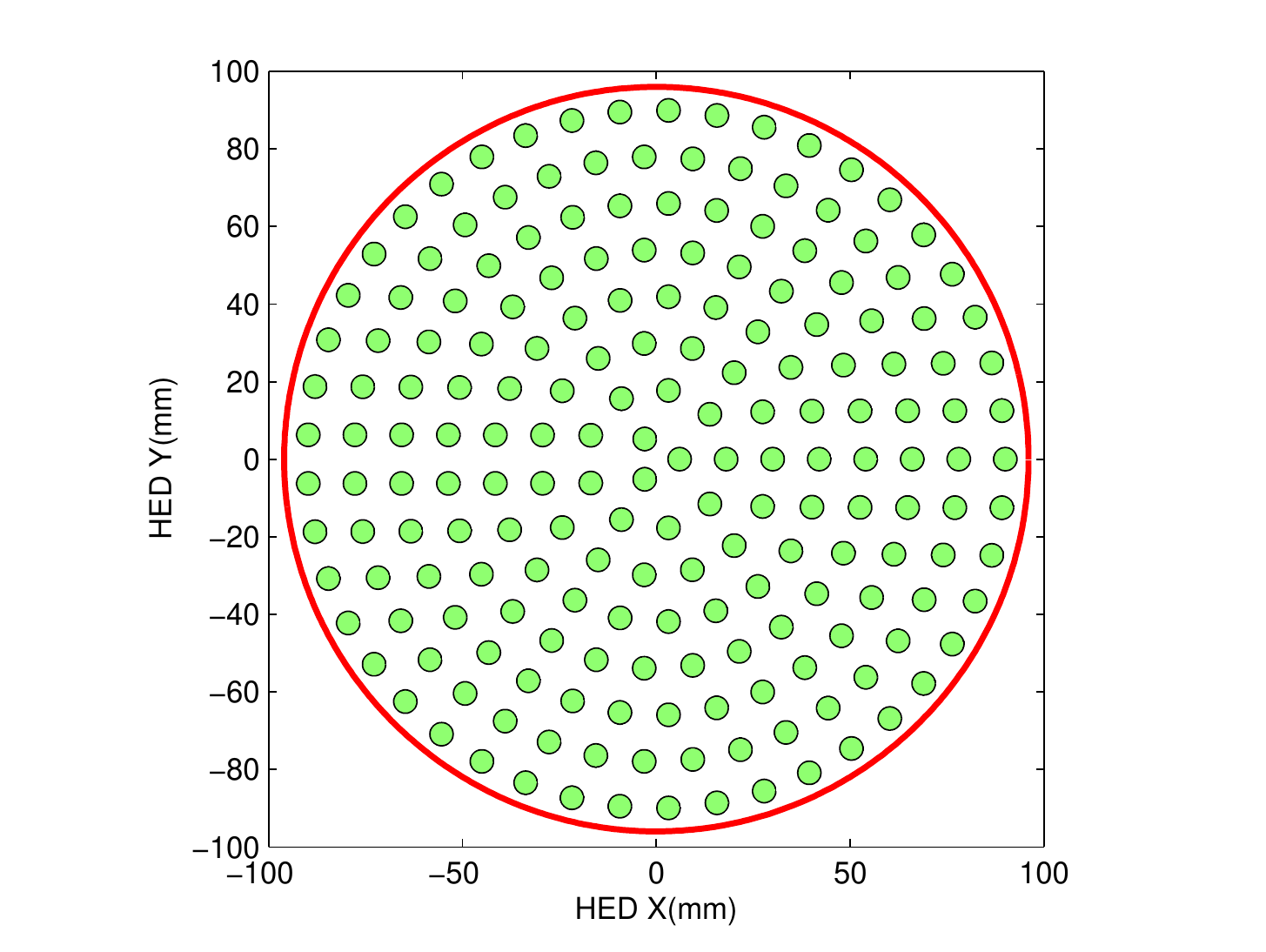}}
  \subfigure[]{
  \includegraphics[width=0.3\textwidth]{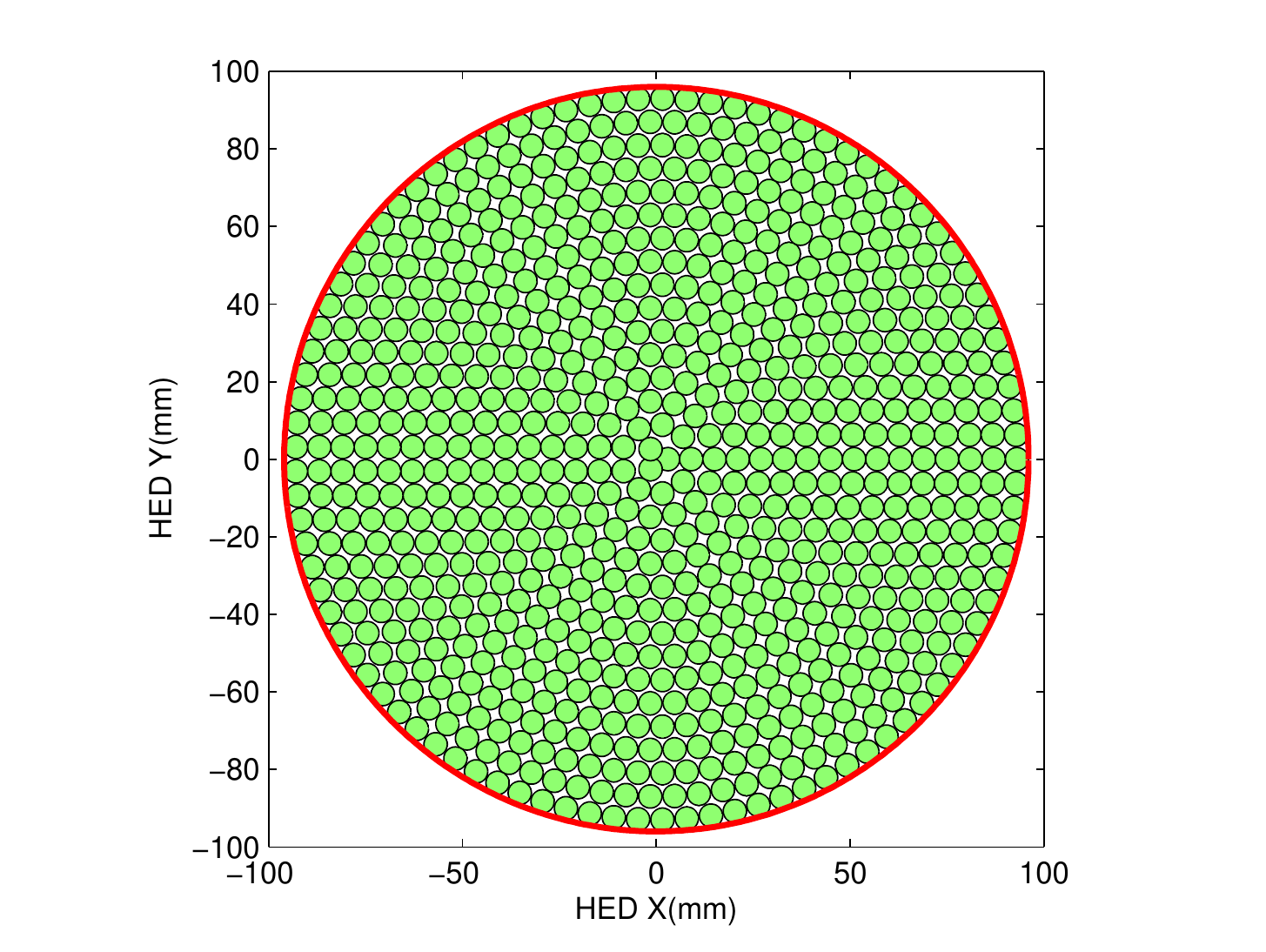}}
  \caption{Distribution of the scan spot over the detector's surface.(a)27 points, (b)192 points, (c)768 points}
  \label{fig:fig3}
\end{figure*}

Because of the detectors' spatial homogeneity, the energy response and the peak efficiency of the detectors as a function of the photon energy were determined by scanning the HED at 27 discrete locations (see Figure \ref{fig:fig3})in X- and $\theta$- direction over the active area while the beam was fixed. During the scan, each position was illuminated by the beam with the same time. During data analysis, the data accumulated at all the 27 points was chosen to form the general response spectrum, however, if the position was shielded by the AGC detector the events at that point would be rejected. In addition, a more detailed scan of the surface area ($\sim$192 or 768 points) was performed at 50\,keV to study the spatial homogeneity of the HED.

The measurements were carried out at 25 different energies, namely from 24 to 30\,keV in 2keV steps, from 31 to 37\,keV in 1keV steps, from 40 to 110\,keV in 10\,keV steps, and at 20, 32.8,130 and 150\,keV. However, in order to avoid the interference of the tungsten characteristic lines, 60\,keV and 70\,keV were replaced by 62\,keV and 72\,keV. Besides, corresponded to the two status of the monochromator, the calibration nearby 50\,keV and 62\,keV were tested twice respectively. There were six different sets of tube voltage/current values for these 25 energies, and in each tube voltage/current state, a background spectrum was recorded (obtained by misorienting the second monochromator crystal, and leaving the monochromator crystals in detuning condition), and used to obtain the net, background subtracted spectrum.

\subsection{Calibration with Radionuclides} \label{sec:3.2 cal with RS}
Another calibration campaign was performed at the NIM radioactivity laboratory using a set of calibrated radioactive sources whose type and properties \citep{NNDC} are list in Table \ref{tab:tab2 nuclides}. Besides, additional background measurements were recorded for long periods shortly before or after the measurements with radioactive sources.

\begin{table}
  \centering
  \caption{Properties of radioactive nuclides used for HED calibration}\label{tab:tab2 nuclides}
    \begin{tabular}{cccccc}
    \hline
    Nuclide & Half-life & Photons & Energy & Intensity \\
     & & & (keV) & (\%)\\
    \hline
  $^{\mathrm139}$Ce & 137.641(20)\,d & $\gamma$ & 165.86 & 80(0) \\
  $^{\mathrm133}$Ba & 10.551(11)\,y & $\gamma$ & 81  & 32.9(3) \\
  {      } & {     }     & $\gamma$ & 356 & 62.05 \\
  $^{\mathrm57}$Co  & 271.74(6)\,d  & $\gamma$ & 122.06 & 85.60(17) \\
  $^{\mathrm241}$Am & 432.6(6)\,y   & $\gamma$ & 59.54 & 35.9(4) \\
  $^{\mathrm109}$Cd & 461.40(12)\,d & $\gamma$ & 88.03 & 3.70(10) \\
    \hline
    \end{tabular}
\end{table}
\begin{figure}[htb]
  \centering
  \subfigure[]{
  \includegraphics[width=0.3\textwidth]{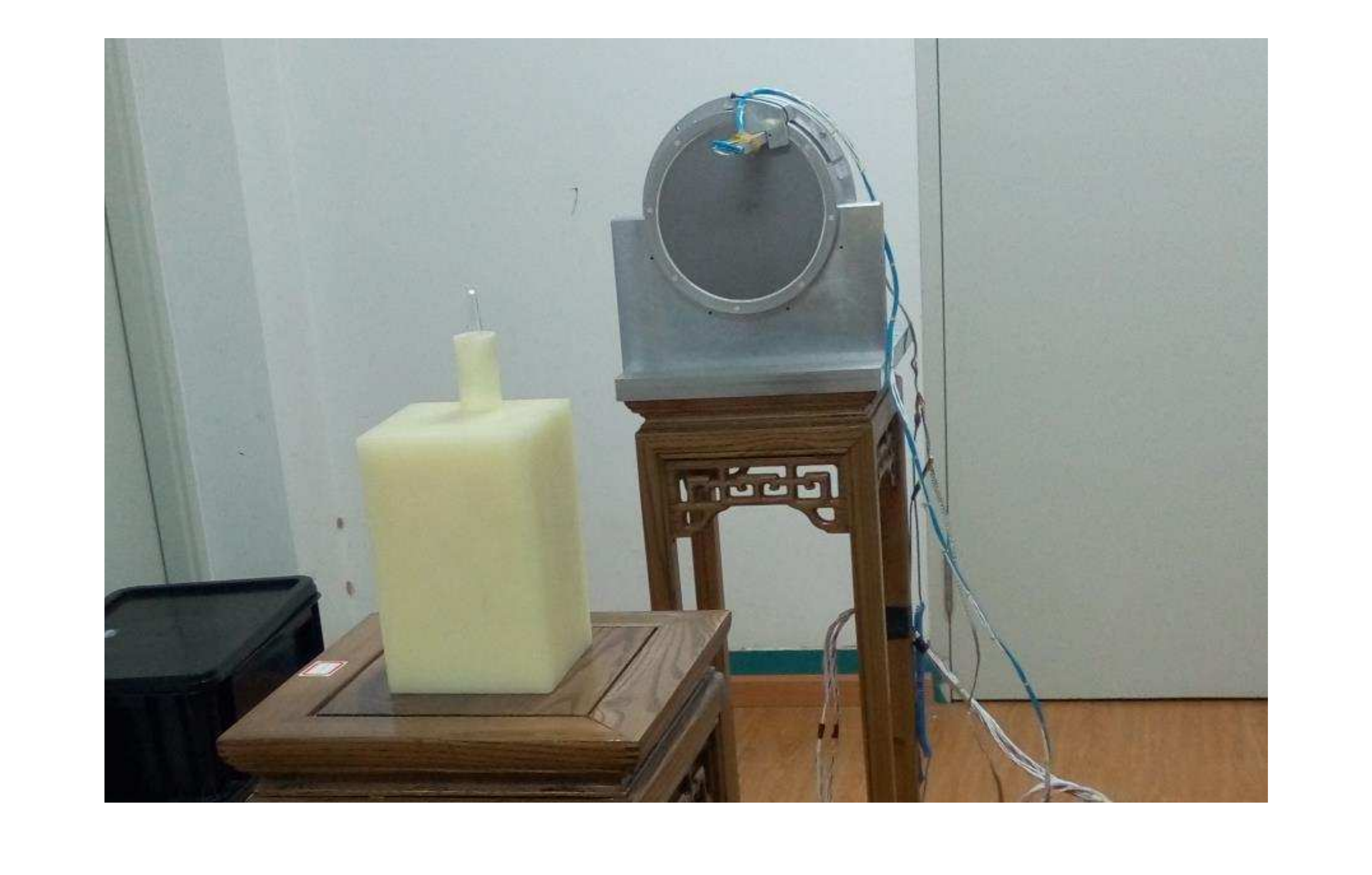}}
  \subfigure[]{
  \includegraphics[width=0.15\textwidth]{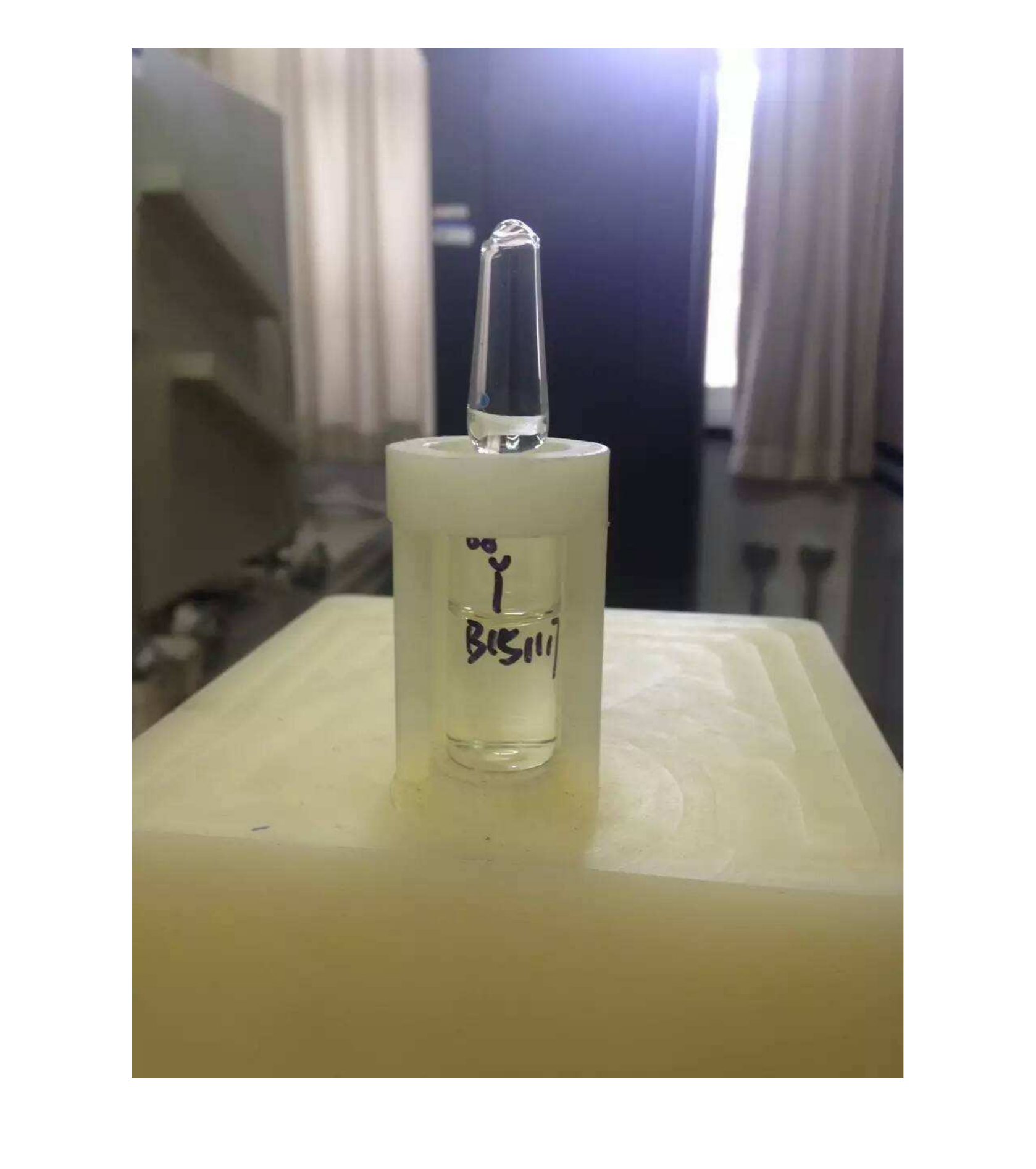}}
  \caption{Calibration campaign with radioactive source at NIM laboratory.(a)Detector and radioactive source holders,(b)Liquid radioactive source}
  \label{fig:fig4}
\end{figure}

The HED and the radioactive sources were fixed on special holders which were placed on wooden stands above the laboratory floor, and the distance between the radioactive sources and the Be-window of the HED was almost 1.5\,m and never changed during measurements (see Figure \ref{fig:fig4}). The lab temperature was also set to $18\pm2{}^{\circ}\mathrm{C}$.

\section{Calibration Data Analysis and Results} \label{sec:4 data analysis}

\subsection{Pulse Shape Discrimination} \label{sec:4.1 PSD}
\begin{figure}[htb]
  \centering
  \includegraphics[width=0.45\textwidth]{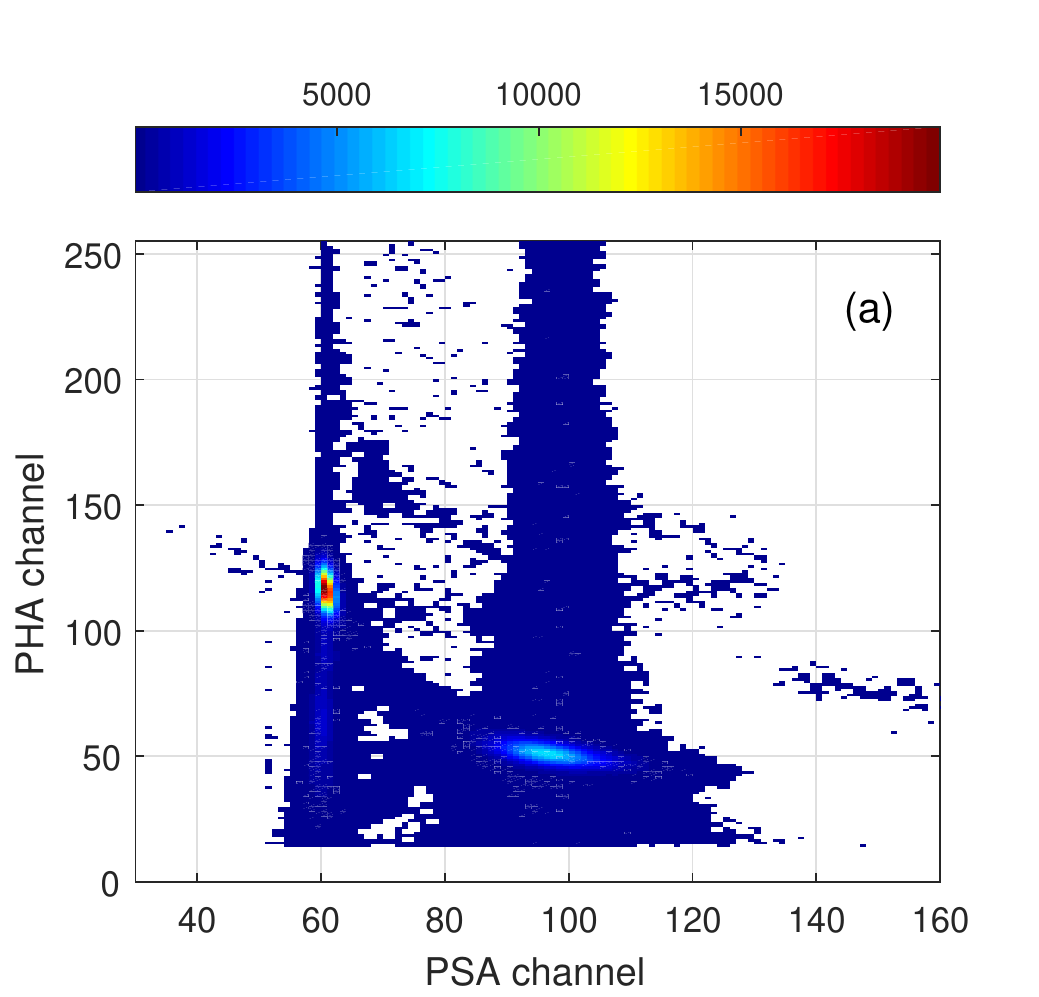}
  \includegraphics[width=0.45\textwidth]{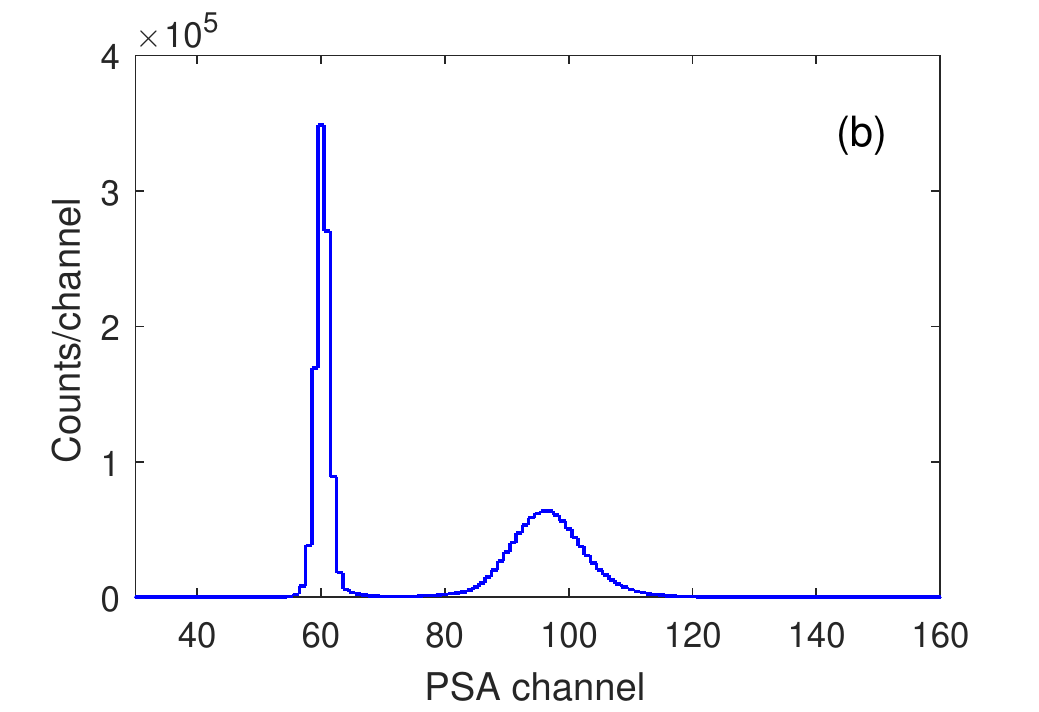}
  \caption{The pulse shape discrimination capability of HE: (a) two-dimensional pulse height-shape histogram measured with monochromatic X-ray radiation at 150keV with HED Z01-25. (b) Integral pulse shape spectrum. The scintillation events detected by each HED can be divided into three classes : NaI events (peak on the left) that deposit energy only in the NaI(Tl) crystal, CsI events (peak on the right) that deposit energy only in the CsI(Na) crystal, and mixed events (bridge connecting the two peaks) that deposit part of their energy in NaI and part in CsI.}\label{fig:fig5 phapsa}
\end{figure}

\begin{figure}[htb]
  \centering
 \includegraphics[width=0.5\textwidth]{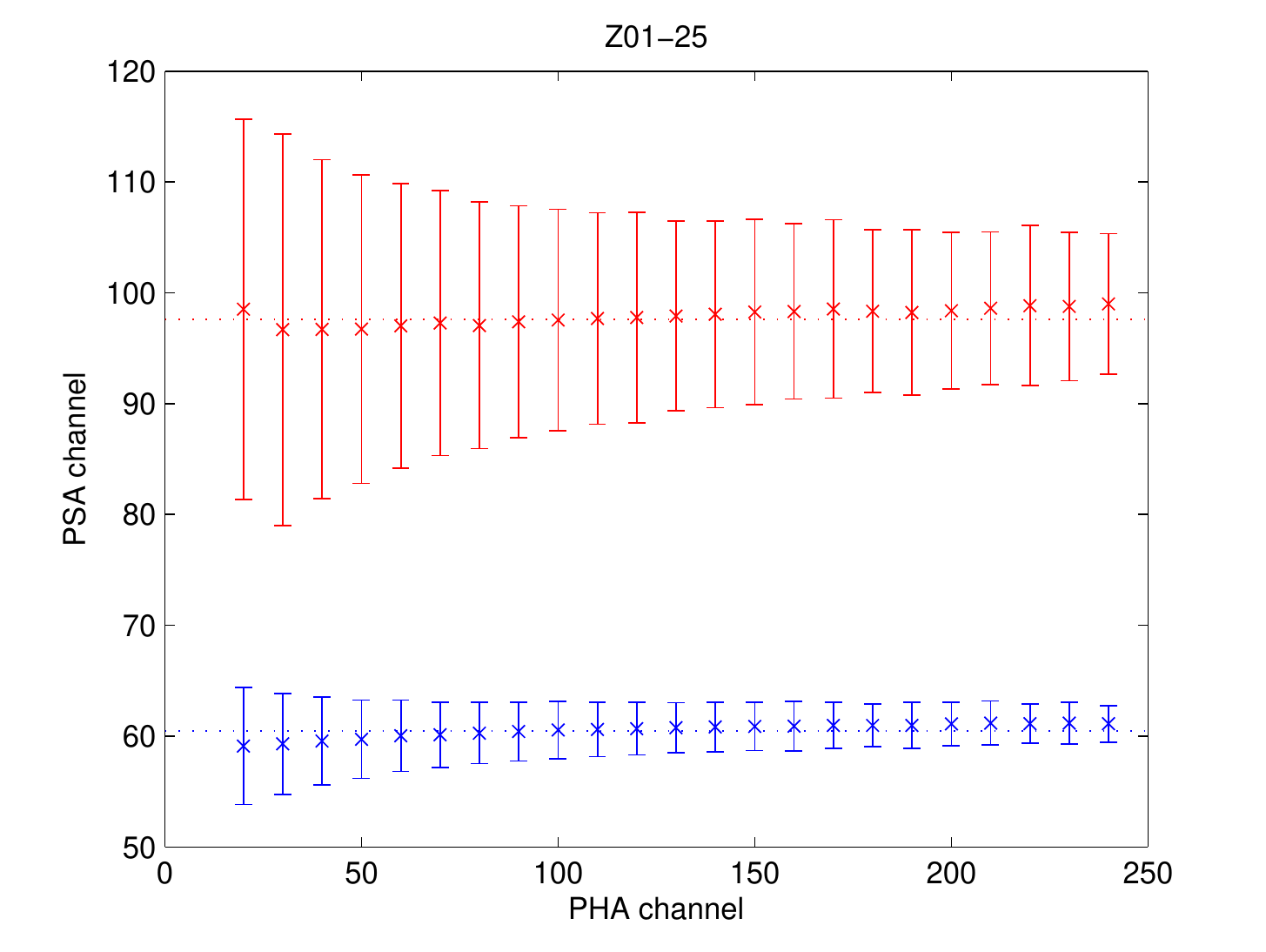}
  \caption{Position of the pulse shape peaks of NaI and CsI crystals as a function of the pulse height channel as derived from background measurements during radioactive sources calibration campaign with HED Z01-25. One error bar is equal to FWHM of each PSA peak.}
  \label{fig:fig6 PSAasPHA}
\end{figure}

Because of the different decay constants of the scintillations produced in the NaI(Tl) and CsI(Na), via Pulse Shape Analyzer (PSA), the events detected by each crystal or both crystals can be discriminated. Figure \ref{fig:fig5 phapsa} shows the PSD capability of HE, it is pretty good. Figure.5(a) is a typical two-dimensional pulse height-shape histogram of the events measured at 150keV monochromatic X-ray with HED Z01-25, and Figure.5(b) is the integral pulse shape spectrum. There are two peaks, the left one is formed by NaI events and the right one is formed by CsI events, the events between the two peaks are mixed events which deposit energy in both crystals. Figure \ref{fig:fig6 PSAasPHA} shows the PSA peak centroid and width (2\,FWHM) of NaI(Tl) and CsI(Na) crystals as a function of the PHA channel which is derived from background data collected during radioactive sources calibration campaign.


 Usually, the PSD figures of merit is quantified as
 \begin{equation}\label{equ:equPSDQ}
   Q=|\frac{PS_1-PS_2}{fwhm_1+fwhm_2}|
 \end{equation}
 where $PS_{1}$ and $PS_{2}$ are the gaussian centriod of the two pulse shape peaks, $fwhm_{1}$ and $fwhm_{2}$ are the FWHM of them. The values $Q$ of all FM HEDs are between  2.6 to 3.0.
 Accordingly, the event with pulse shape between 55 and 70 channels is identified as NaI event during our data analysis. And in this paper we just report the NaI calibration results.

 \subsection{Processing of Calibration Runs} \label{sec:4.2 peak fit}
A series net, background subtracted spectra collected with HED/NaI Z01-25 are shown below.  The six spectra in Figure \ref{fig:fig7 BKG-sub spectra 1}, which measured at different energies of monochromatic X-ray irradiation (MR), highlight the appearance of an important feature of the NaI spectra. Below the Iodine K-edge energy of 33.17\,keV, spectra display only the full-energy peak. However, because of the threshold of PHA, the escape peak of Iodine(I) cannot be seen for photons of energy higher than the K-edge energy until the energy approach to 44keV nearby (the escape peak energy has exceeded the threshold energy). With the photon energy increasing, the escape peak appears to the left of the full-energy peak, whose energy equals the energy of the X-ray photon from the monochromator minus the energy of the escaped X-ray. The contributions of the different Iodine fluorescence lines can not be distinguished by HED, because its energy resolution is not good enough. The peak area ratio between the escape peak and the full-energy peak become lower with increasing photon energy. In the spectra with monochromator set to 130\,keV and 150\,keV the escape peak becomes not very obvious and submerged in the full-energy peak because of the low probability and large beam spread. The full-energy peak was fitted with a single Gaussian shaped curve, besides, for energies between 50\,keV and 110\,keV, the Iodine escape peak is also fitted with a single Gaussian function. A program computed the fit, extracted the fitting parameters (such as the peak area, the peak centroid and the FWHM) and computed the errors.

NaI spectra from radioactive sources(RS) are shown in Figure \ref{fig:fig8 BKG-sub spectra 2}. The spectra of the lab sources are more complicated and show more features by such effects as Compton scattering internal to the source. In fitting full-energy peak, some other functions (linear, quadratic or Gaussian) had added to the main Gaussian in order to account for non-photo-peak contributions \citep{Fermi..GBM..2009}.
 \begin{figure}[htb]
  \centering
  \includegraphics[width=0.5\textwidth]{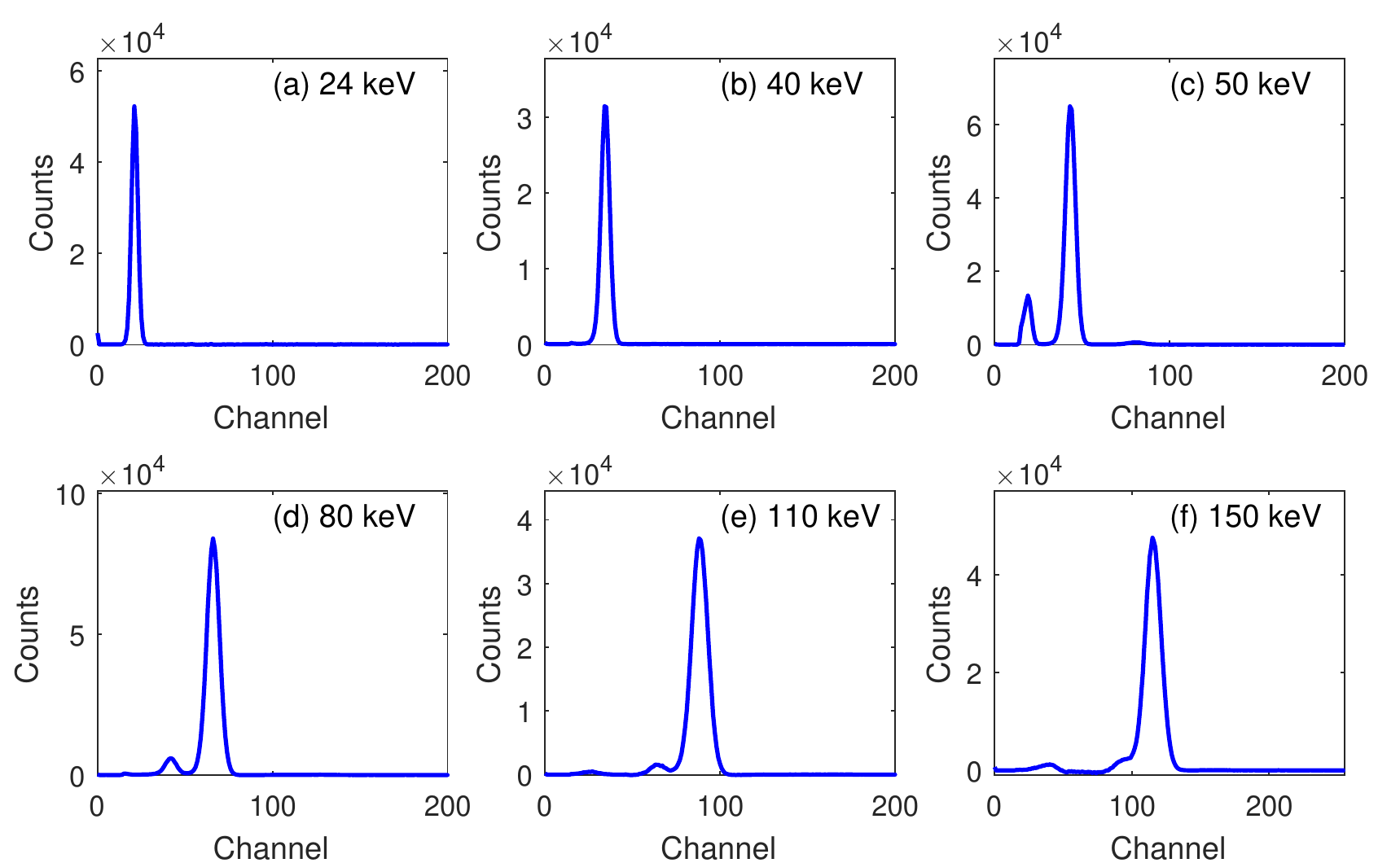}
  \caption{Normalized background-subtracted spectra measured at HXCF with HED Z01-25. Results for six different photon energies are shown: (a)24\,keV,(b)40\,keV,(c)50\,keV,(d)80\,keV,(e)130\,keV and (f)150\,keV}\label{fig:fig7 BKG-sub spectra 1}
\end{figure}
\begin{figure}[!htb]
  \centering
  \includegraphics[width=0.5\textwidth]{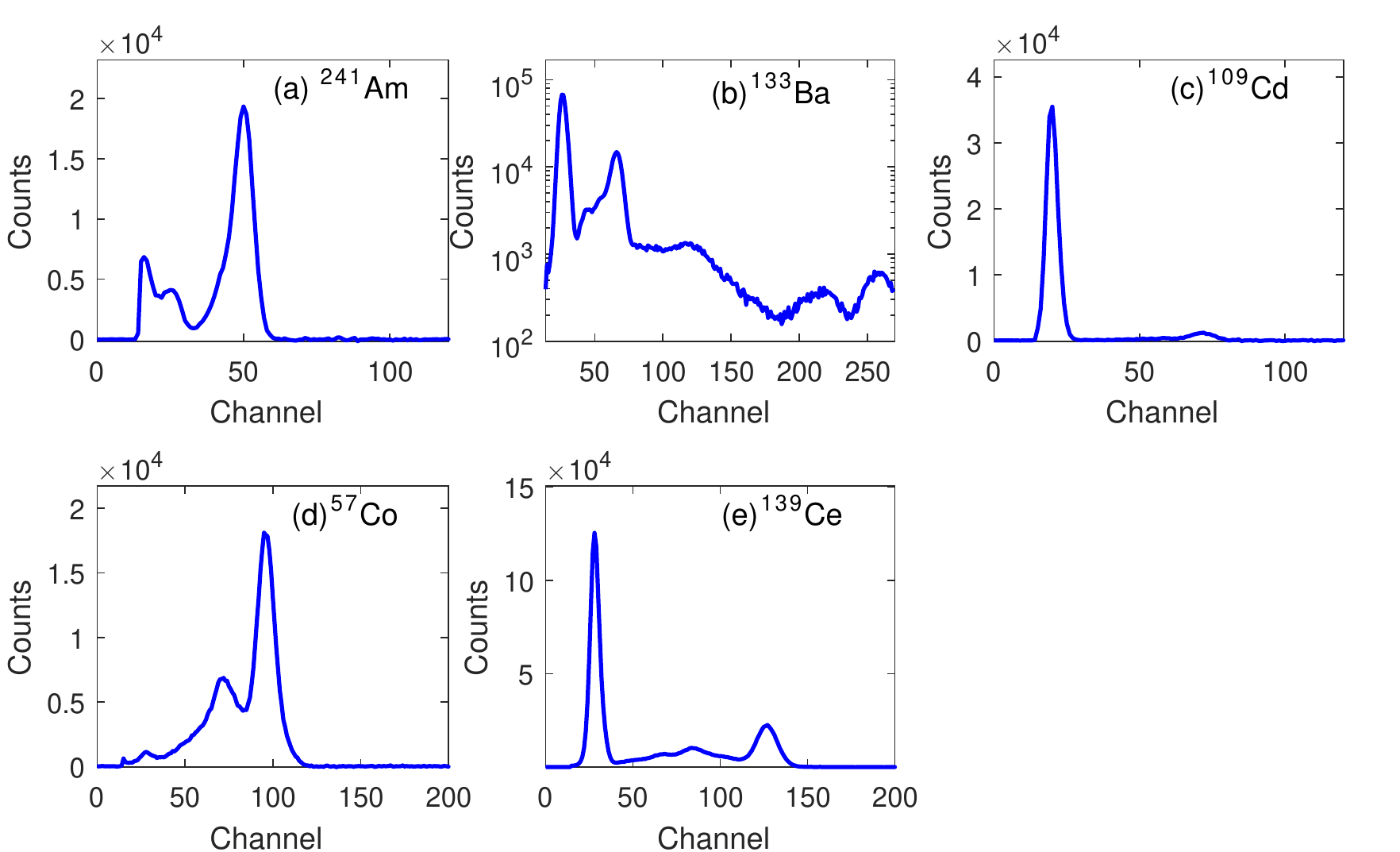}
  \caption{Normalized background-subtracted spectra of HED Z01-25 irradiated by radioactive sources (RS): (a)\,$^{\mathrm241}$Am,(b)\,$^{\mathrm133}$Ba,(c)\,$^{\mathrm109}$Cd,(d)\,$^{\mathrm57}$Co and (e)\,$^{\mathrm139}$Ce.}
  \label{fig:fig8 BKG-sub spectra 2}
\end{figure}

\begin{figure*}[htb]
   \centering
   \subfigure{
   \includegraphics[width=0.3\textwidth]{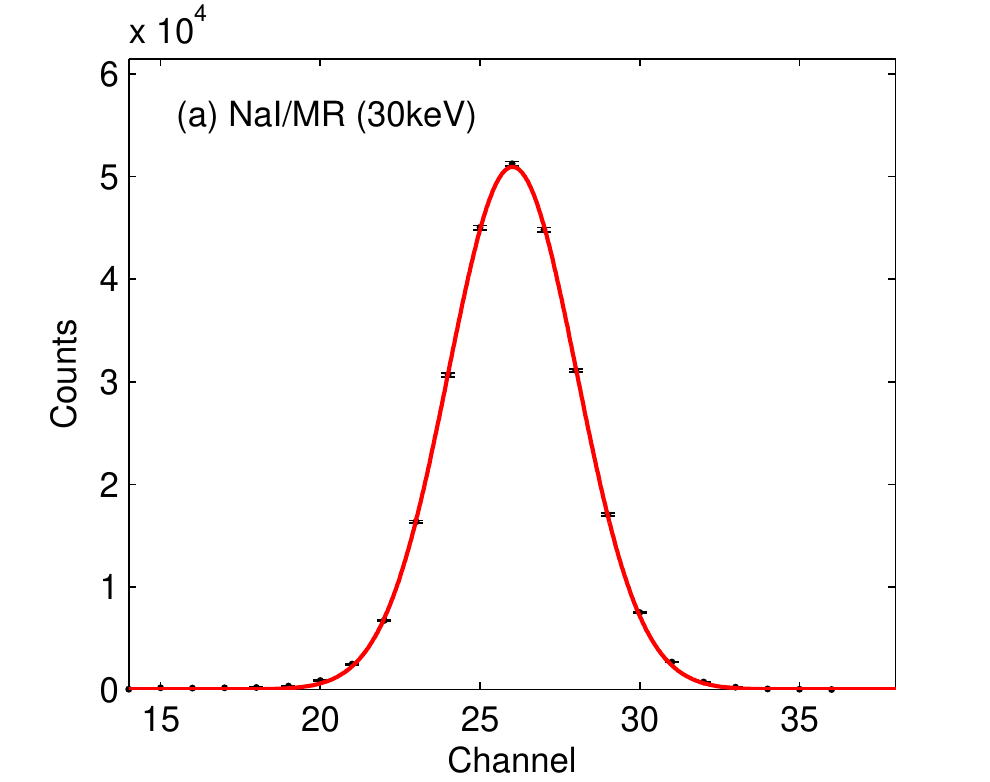}}
   \subfigure{
   \includegraphics[width=0.3\textwidth]{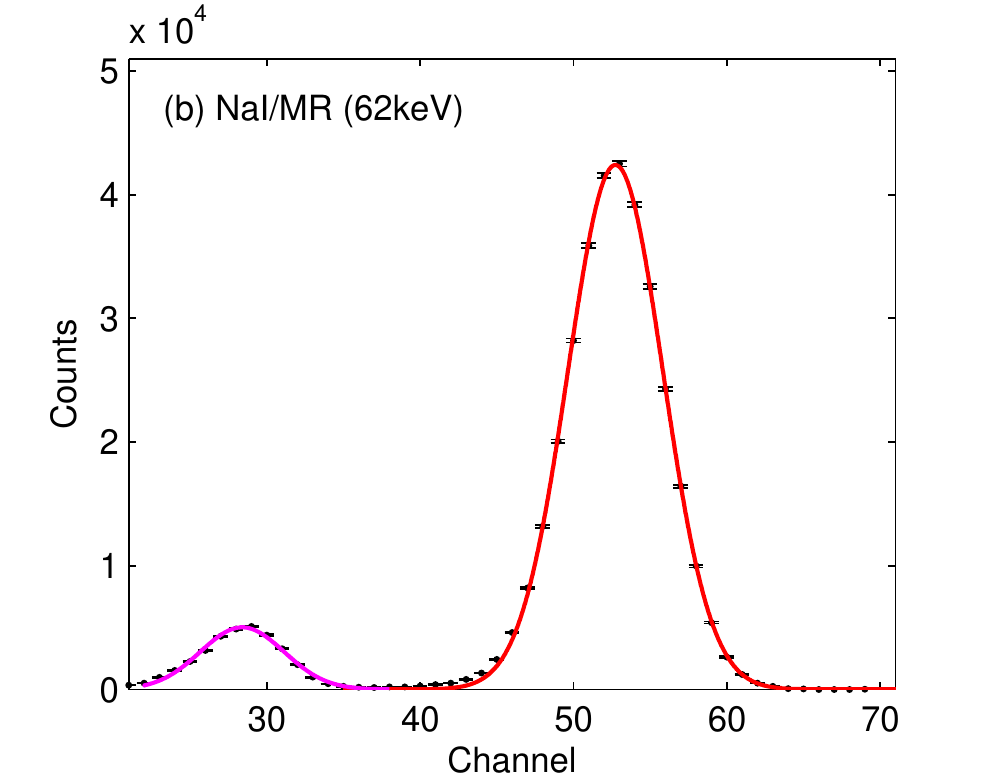}}
   \subfigure{
   \includegraphics[width=0.3\textwidth]{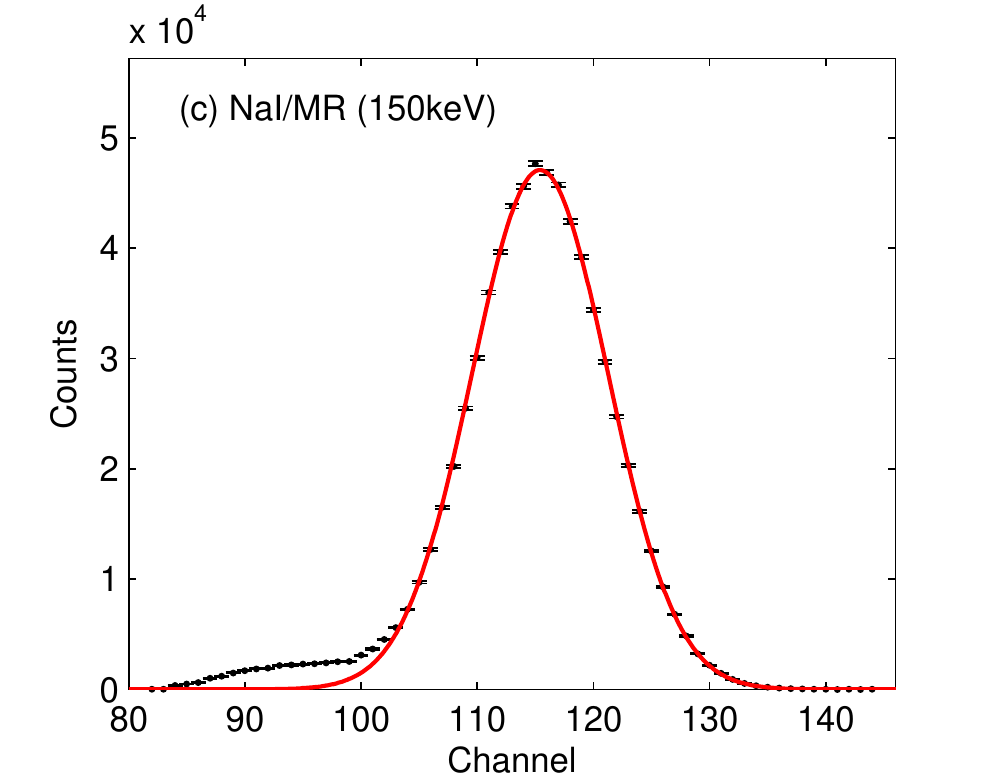}}
   \subfigure{
   \includegraphics[width=0.3\textwidth]{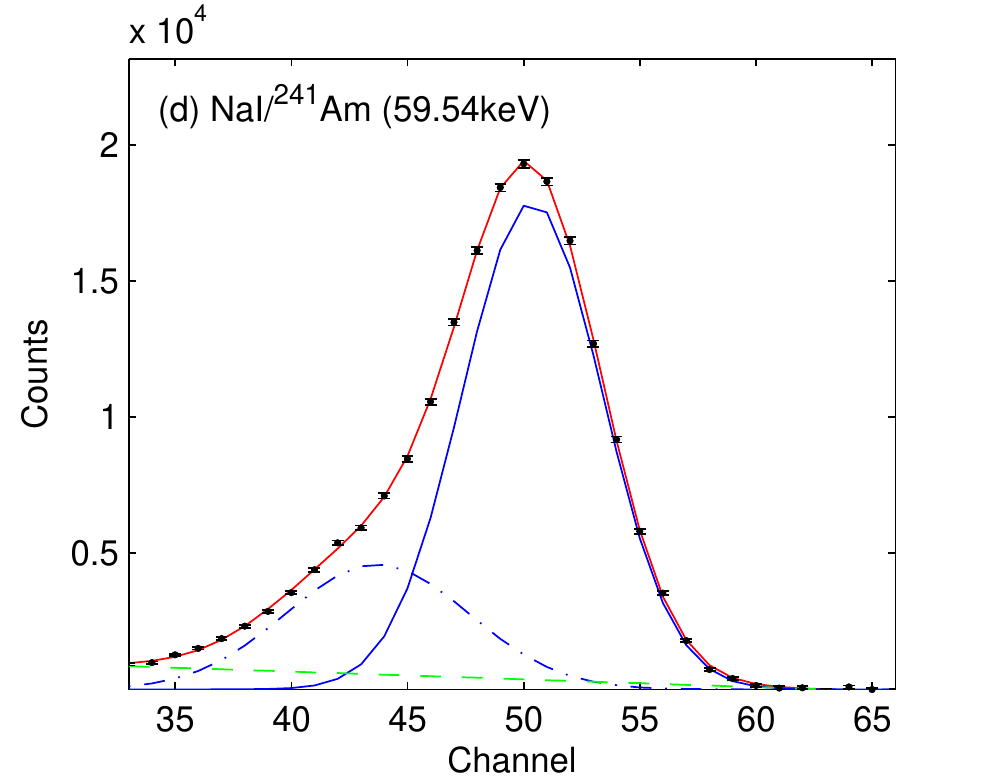}}
   \subfigure{
   \includegraphics[width=0.3\textwidth]{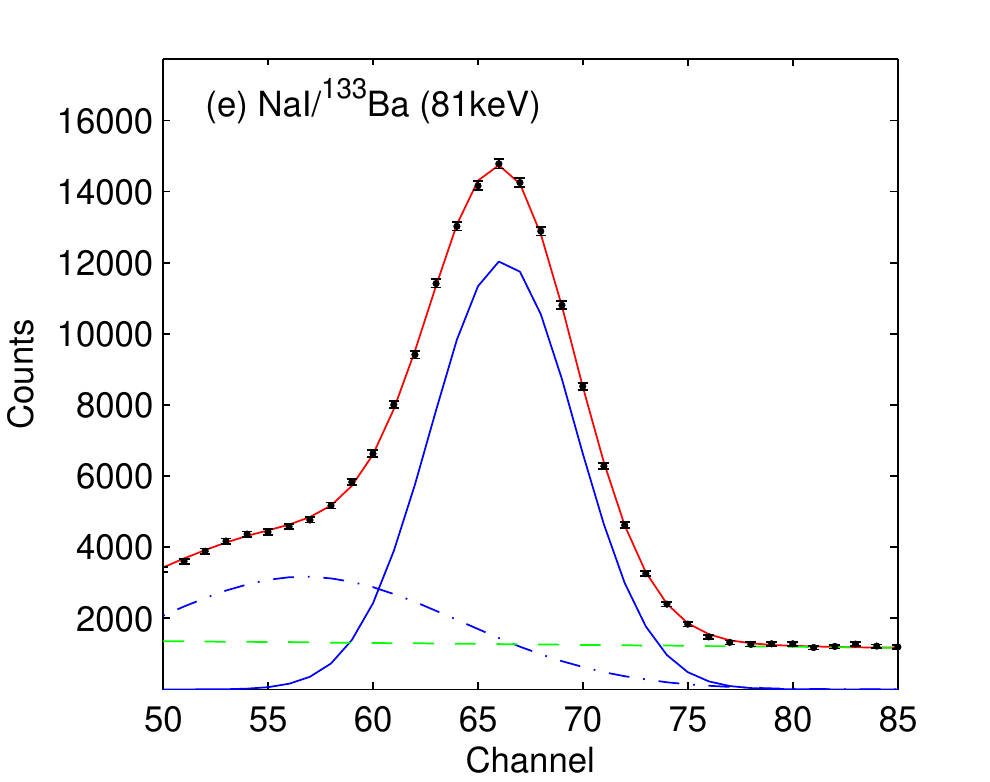}}
   \subfigure{
   \includegraphics[width=0.3\textwidth]{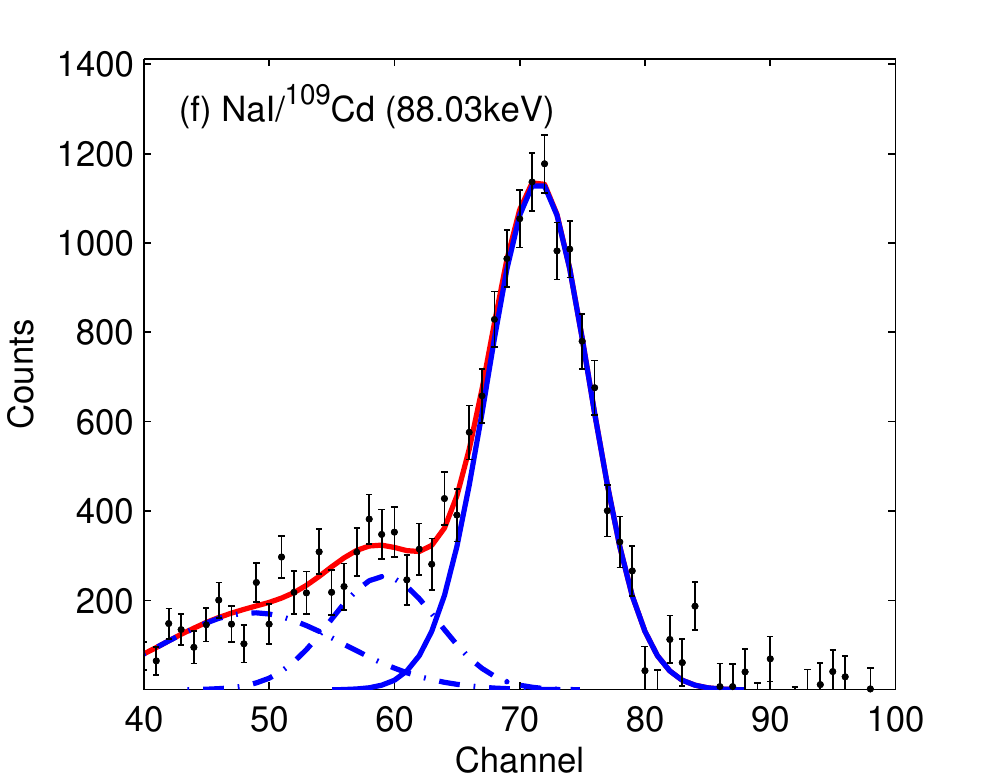}}
   \subfigure{
   \includegraphics[width=0.3\textwidth]{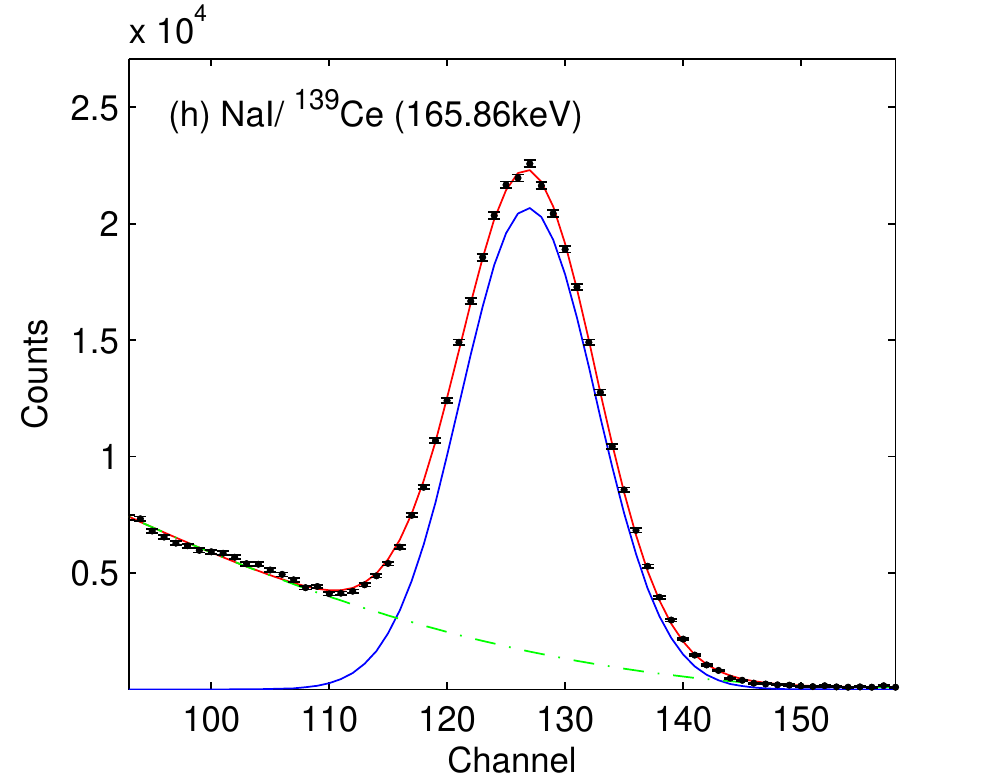}}
   \subfigure{
   \includegraphics[width=0.3\textwidth]{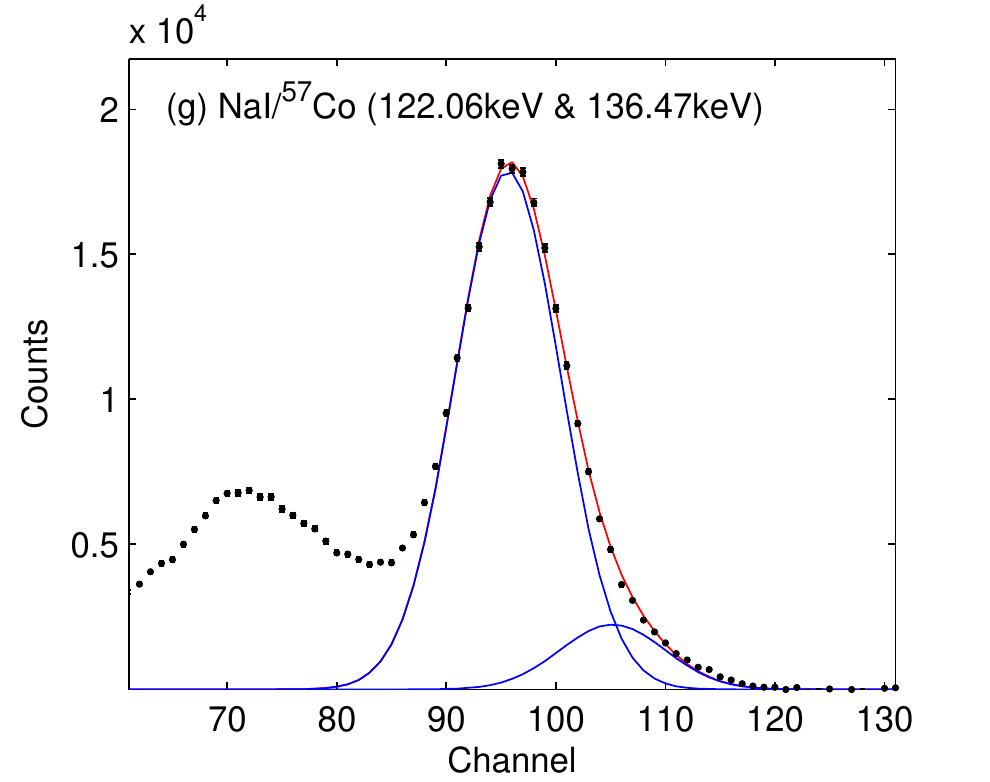}}
   \subfigure{
   \includegraphics[width=0.3\textwidth]{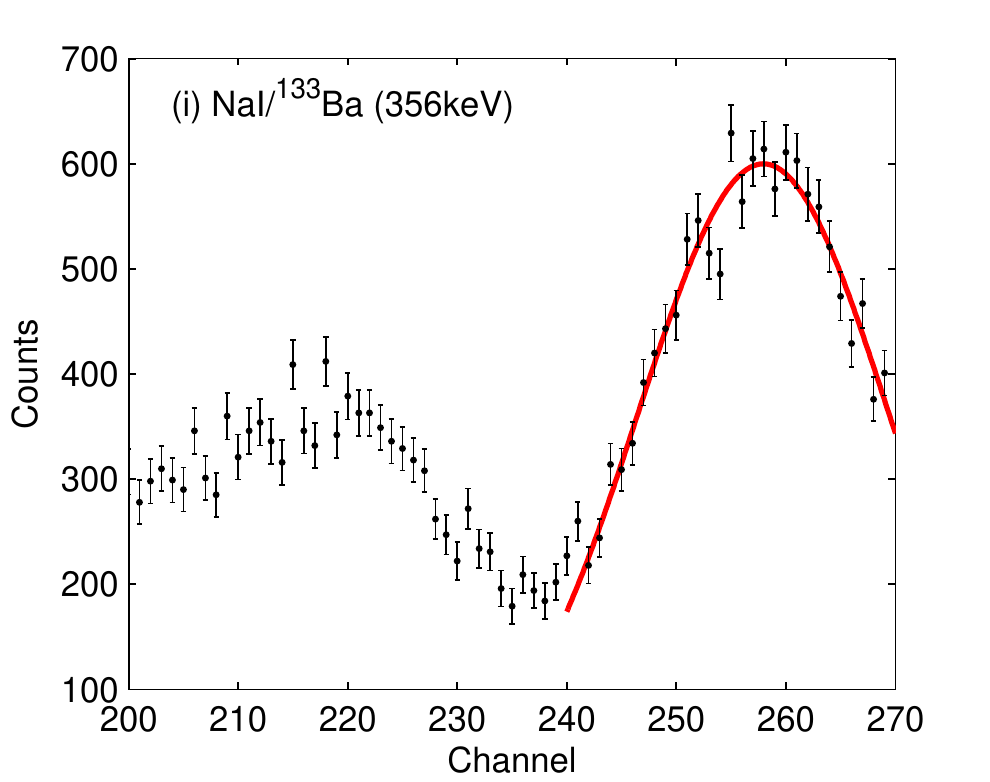}}
\caption{Full-energy peak analysis of HED/NaI Z01-25. Data points (in black) are plotted with statistical errors. Line fits(solid red curves) arise from the superposition of different components: (i) one(or more) Gaussian functions describing the full-energy peak(s) (solid blue curves);(ii) secondary Gaussian function modeling the Iodine escape peaks (solid magenta curve);(iii)one (or more) Gaussian function modeling  other unknown background(dash-dot blue curves);(iv)a constant, linear or quadratic function accounting for background contributions (dotted green curves). For MR (panel a-c) , RS $^{\mathrm57}$Co (panel g) and $^{\mathrm133}$Ba 356\,keV (panel i) line analysis, no background was modeled. }
\label{fig:fig9 peak fit}
\end{figure*}

The Figure \ref{fig:fig9 peak fit} shows some fitting results for HED/NaI Z01-25. For MR line analysis, the Gaussian components, describing the full-energy peaks and escape peak are shown as solid red and solid magenta curves, respectively. For RS lines, fits to the data are shown in red. One Gaussian function describing the full-energy peak is shown as solid blue curve. Dashed green curves represent background and dash-dot blue lines modeling some unknown background features. For $^{\mathrm133}$Ba 356\,keV gamma-ray line analysis, a single Gaussian was performed to model the full-energy peak.

\subsection{Channel-Energy Relation} \label{sec:4.3 E-C}
\begin{figure}[!htb]
  \centering
  \includegraphics[width=0.5\textwidth]{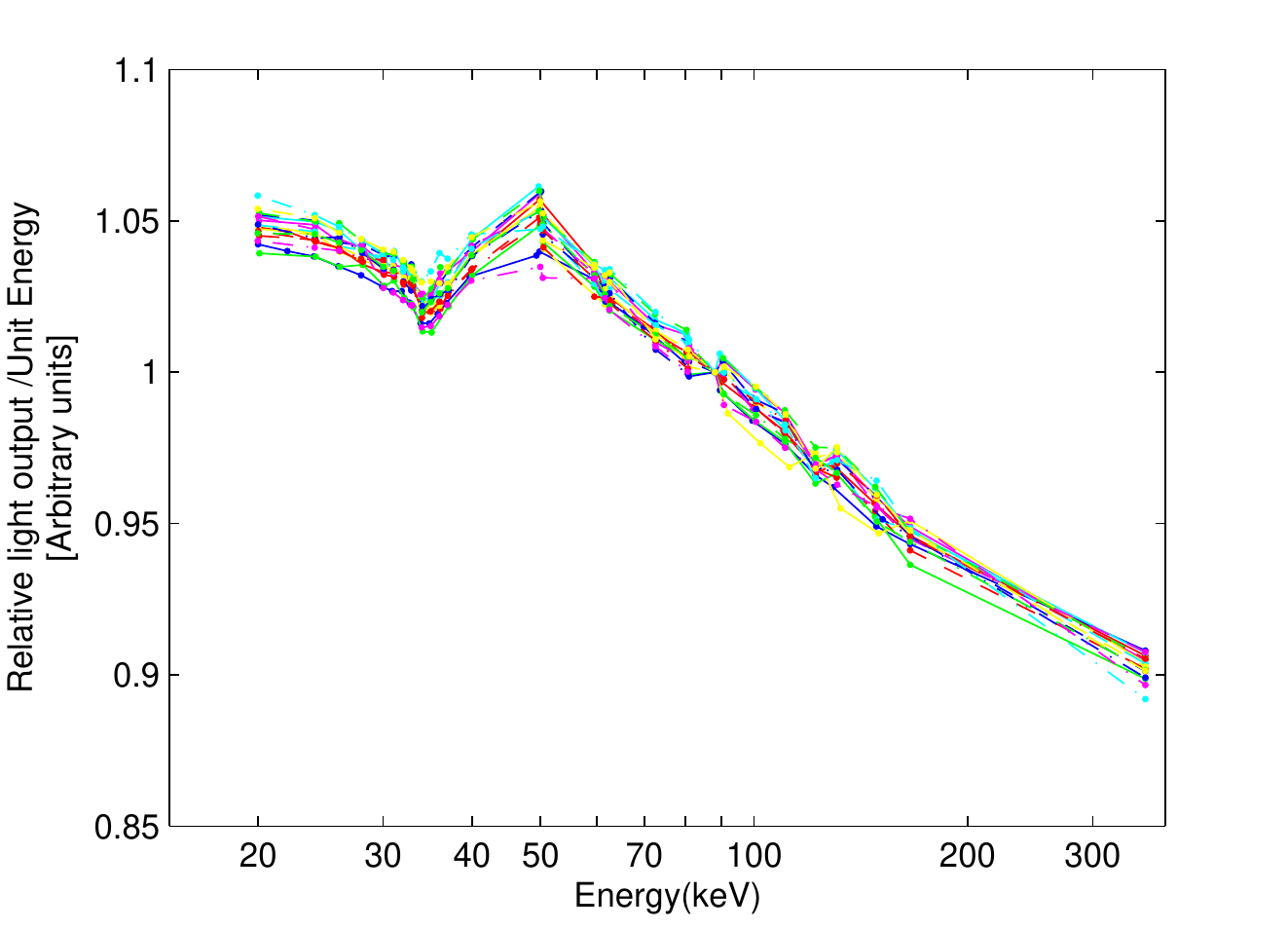}
  \caption{The Photopeak-\emph{nPR} of each FM HED as a function of photon energy, normalized to unity at 88.03keV.}
  \label{fig:fig10 NaI nPR}
\end{figure}

\begin{figure}[!htb]
  \centering
   \includegraphics[width=0.5\textwidth]{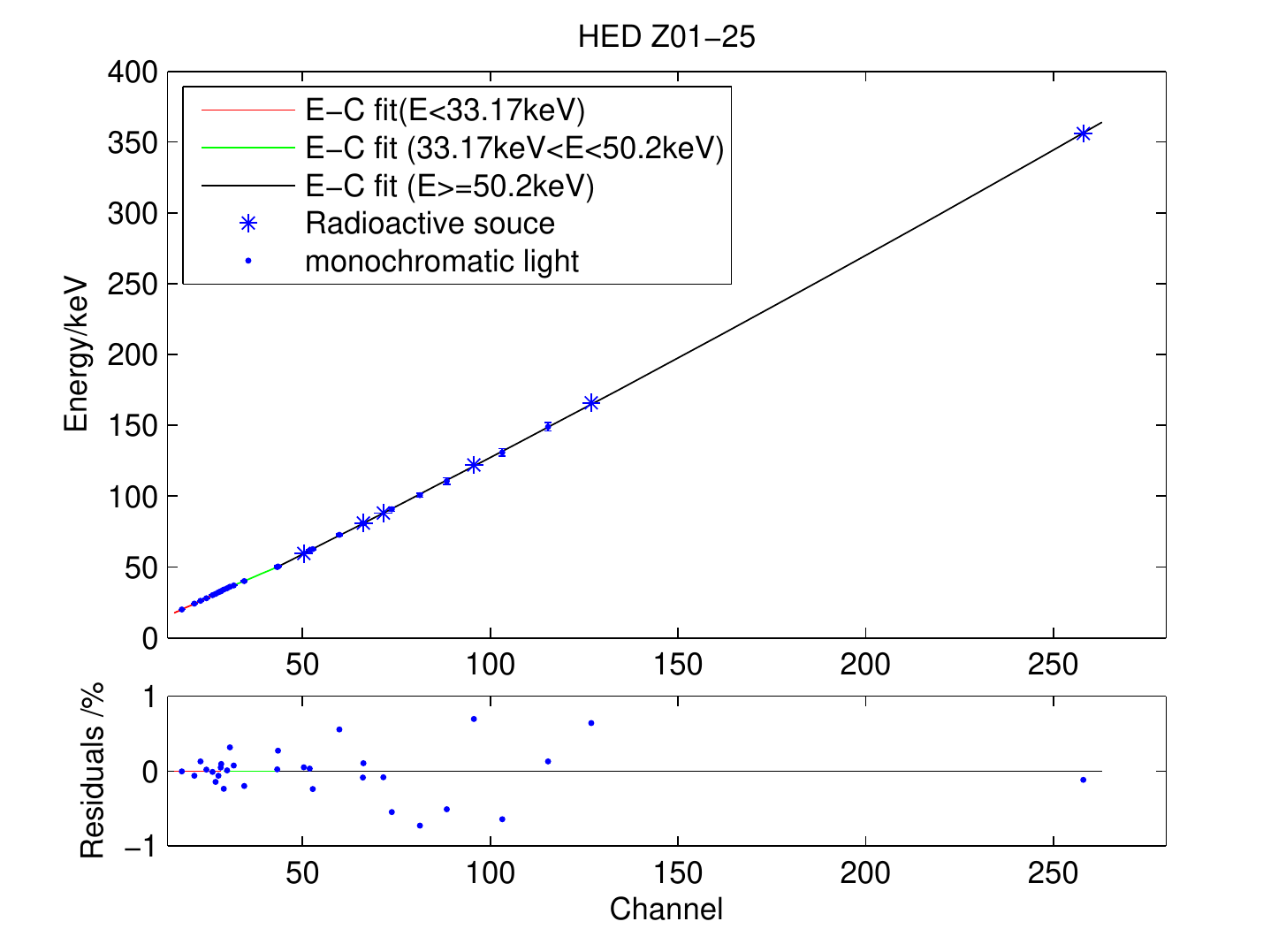}
  \caption{Channel-energy relations for HED/NaI Z01-25.}\label{fig:fig11 E-C}
\end{figure}

Many studies have indicated that the light output response of NaI(Tl) to X/$\gamma$-rays is non-proportional (\citep{NaI..nRP..1998},\citep{Khodyuk..nPR..2010}). Such non-proportional response (\emph{nPR}) must be taken into account when relating the pulse height channel to X/$\gamma$-ray energies. In this work the \emph{nPR} of NaI(Tl) at energy $E_X$ was defined as the light output divided by X-ray energy, and the light output equals the peak centroid channel ($pch$) (obtained from the fitting process) minus the step of our electronics (equals 1 channel, which was tested before and after the calibration campaigns). Figure \ref{fig:fig10 NaI nPR} shows the \emph{nPR} of each flight HED/NaI(Tl) crystal as a function of photon energy, normalized to 88.03\,keV line from $^{\mathrm109}$Cd. The data points include both measurements taken with monochromator and the radionuclides. The curves are similar to results reported before, a clear dip in the plot at the K-shell binding energy in Iodine is observed, a decrease with increase in X-ray energy from 20-33.17\,keV followed by an increase in the range 33.17-50\,keV, then the \emph{nPR} decrease with increase in $E_X$. Fortunately, the spectra of the monochromatic X-ray at 50\,keV was measured twice corresponding to the different monochromator statuses, and there are 0.3-0.5keV difference between the two beam energy(determined by the LEGe). Based on the data of all detectors, we set the inflection point at 50.2\,keV. Therefore the energy rang was split into three regions to compute the channel-energy relation, one below the K-edge energy, one from the K-edge energy to 50.2keV, and one above the 50.2keV. Data measured with the radioactive sources and at HXCF were fitted together with quadratic function for each energy range. Figure \ref{fig:fig11 E-C} shows an example of the channel-energy relation calculated for HED/NaI Z01-25. The analysis results of all detectors are similar. Almost all calculated relations give fit residuals less than 1\%.

After directly applying the channel-energy conversion, we found that the energy intervals between full-energy peak and escape peak increases with the rising X-ray energy from 50 to 110\,keV, which is against physics principle. But before conversion the channel intervals ($\Delta\,ch$) between the two peaks remain unchanged for different X-rays.  Finally we found that this phenomenon is caused by the \emph{nPR} of NaI(Tl) to electrons, so the E-C relation of the full-energy peak is different to that of escape peak. Then the energy of the escape peak can be given by the following expression:
 \begin{equation}\label{equ:equ1}
   E_{escape}=E(pch)-E(\Delta\,ch+step)
,\end{equation}
where $E(pch)$  is the photopeak energy converted using the relations above,$E(\Delta\,ch+step)$ is calculated using the channel-energy relation of E < 33.17\,keV, which equivalent to the mean value of the escape energies of Iodine K X-ray fluorescence (nearly 29.2\,keV). This phenomenon is also simulated in our detector response simulation.
\subsection{Energy Resolution} \label{sec:4.4 resolution}
\begin{figure}[!htb]
  \centering
  \includegraphics[width=0.5\textwidth]{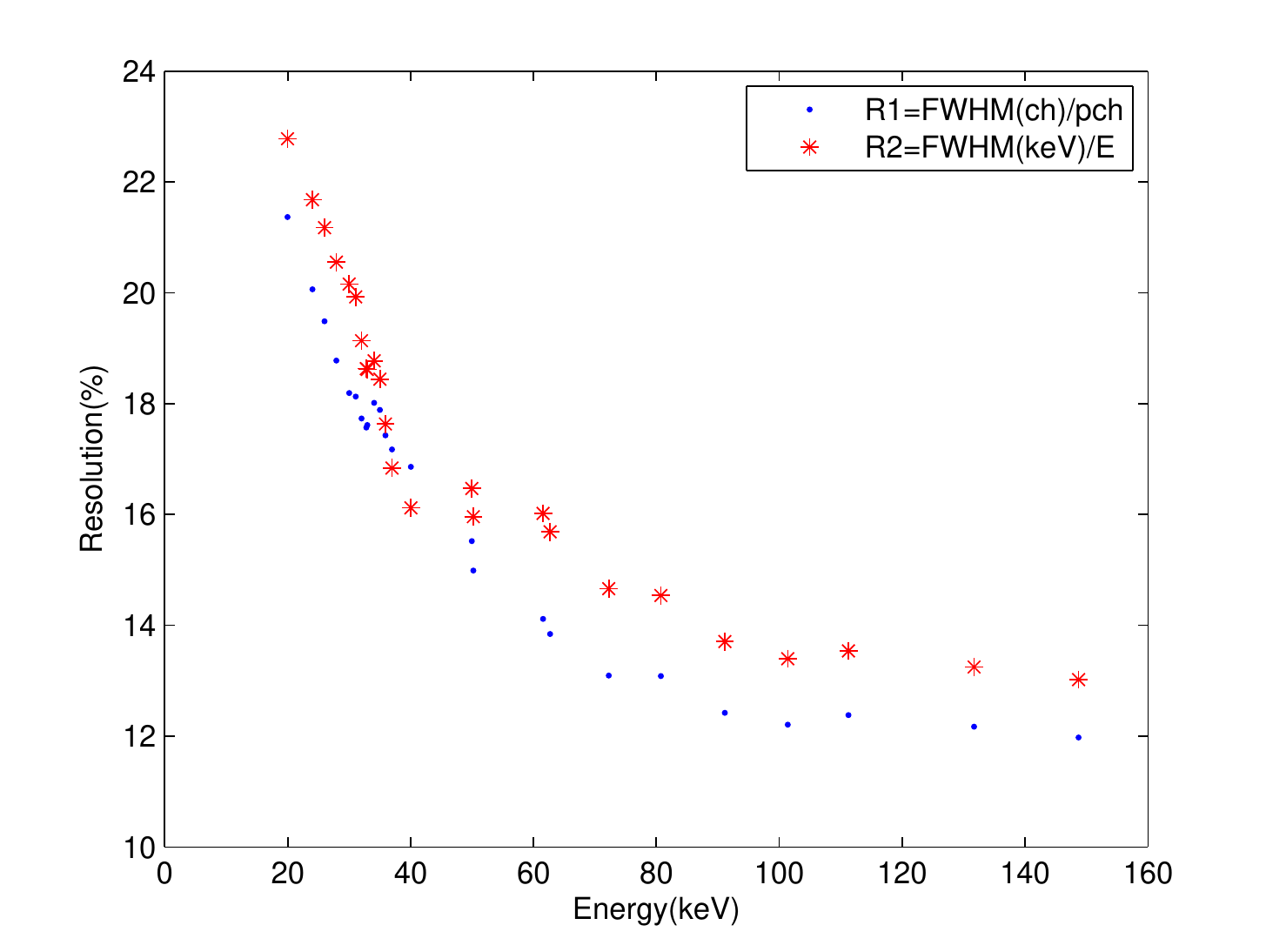}
  \caption{Resolution in two definitions for the monochromatic X-ray irradiation recorded with the HED/NaI Z01-25}
  \label{fig:fig12 resolution}
\end{figure}

The energy resolution of a detector can be conventionally defined as the FWHM divided by the location of the peak centroid $pch$ (\citep{Knoll}). However, it can also be defined as the FWHM in keV divided by the energy of the X-ray. In this paper the former definition is marked as $R_1$  , and the latter is $R_2$ . The $R_2$  for the monochromator X-ray irradiation was derived by refitting the peaks using the same spectra and the channel-energy relationship results previously obtained. For the spectra measured at HXCF with the HED/NaI, both definitions of resolution are plotted in Figure \ref{fig:fig12 resolution} versus $E_X$. There are some difference between $R_1$ and $R_2$, because the response of the NaI(Tl) is non-linearity, after convert the measured spectra in channels to energies by directly applying to detector the corresponding channel-energy relation previously obtained, the shape of the full-energy peak will be distorted. So the definition $R_1$  was used in the analysis.

\begin{figure}[!htb]
  \centering
 \includegraphics[width=0.5\textwidth]{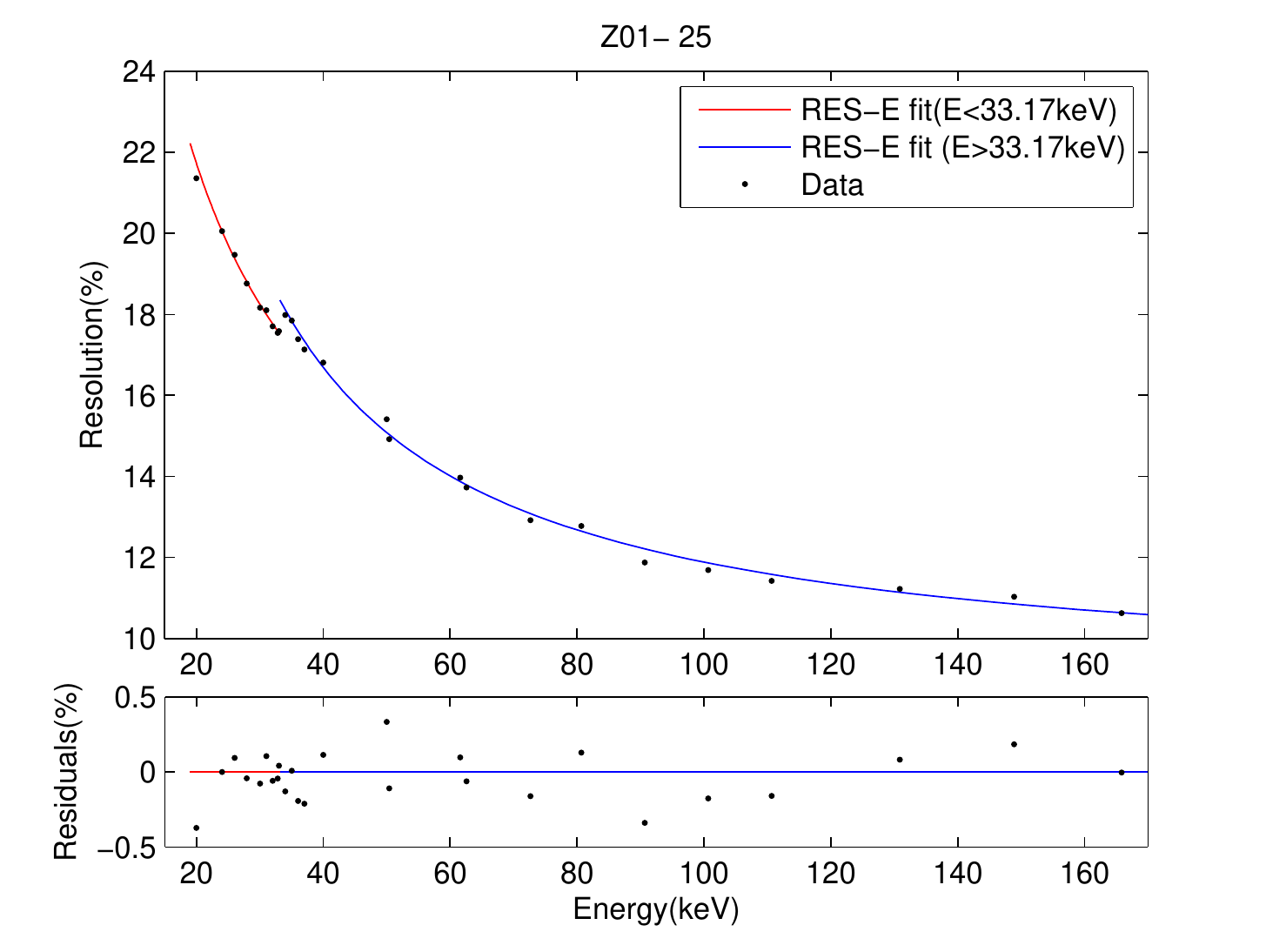}
  \caption{The energy resolution of the HED/NaI Z01-25 as a function of energy.}\label{fig:fig13_ResAsE}
\end{figure}

The relative intrinsic energy width of the monochromatic X-ray beam is $\eta$ (\%), which is determined by the LEGe detector, then the intrinsic resolution $R$ of the HED can be calculated as $R=\sqrt{{R_1}^2-\eta^2}$  . The $R$ decreases with the X-ray energy and a clear steplike change can be found at energy around K-edge. As a result two different fits were performed below and above the iodine K-edge energy, and the following function was adopted for both regions:
\begin{equation}\label{equ:equ2}
  Res(E)={\frac{\sqrt{a^{2}+b^{2}E+c^{2}E^{2}}}{E}}\times{100\%}
\end{equation}
Because the resolution calibrated with the lines of radioactive sources have greater uncertainties, only the 165\,keV from $^{\mathrm139}$Ce was used in fitting. Fit  result for HED/NaI Z01-25 is show in Figure \ref{fig:fig13_ResAsE}. For all detectors, the calculated relations give fit residuals below 0.5\%, and the FM Z01-24 show poorer energy resolution compared to the others. At 60\,keV, the energy resolution of FM Z01-24 is 14.9\%, and that of the other FM HEDs is better than 14\%. Finally, for the actual detector response at discrete energies, the FWHM in keV would be calculated by the Resolution-energy relationship and the channel-energy relationship obtained above, which will be applied to the simulation.

The dependence of the energy resolution on High Voltage\,(HV) power supply of phoswith PMT was investigated during the PMT specification test. Results show that the energy resolution remains constant when the HV is between 920 to 1200V\,(nominal HV is around 1000V).

\subsection{Intrinsic Peak Efficiency} \label{sec:4.5 QDE}
The intrinsic peak efficiencies (full-energy peak efficiency) (\citep{Knoll}) of HED were determined through the calibration campaigns performed at HXCF and computed as:
\begin{equation}\label{equ:equ3}
  \varepsilon_{inp}(E_X)=\frac{n(E_X)}{I(E_X)\kappa_I}
  =\frac{n(E_X)\cdot\varepsilon_{LEGe}(E_X)}{n_{LEGe}(E_X)\cdot\kappa_I}
\end{equation}

\begin{figure}[!htb]
  \centering
  \subfigure{
  \includegraphics[width=0.45\textwidth]{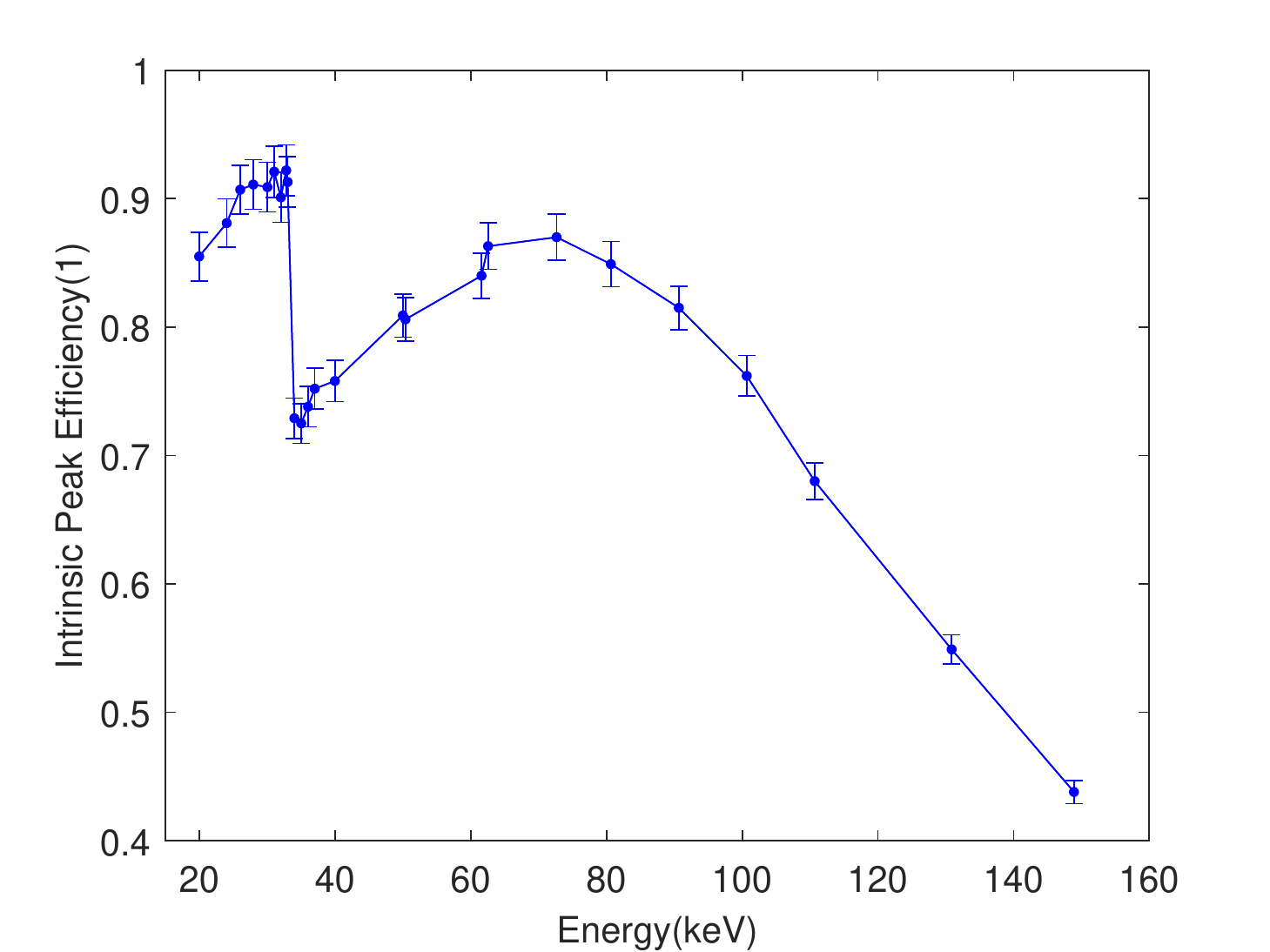}}
  \subfigure{
  \includegraphics[width=0.45\textwidth]{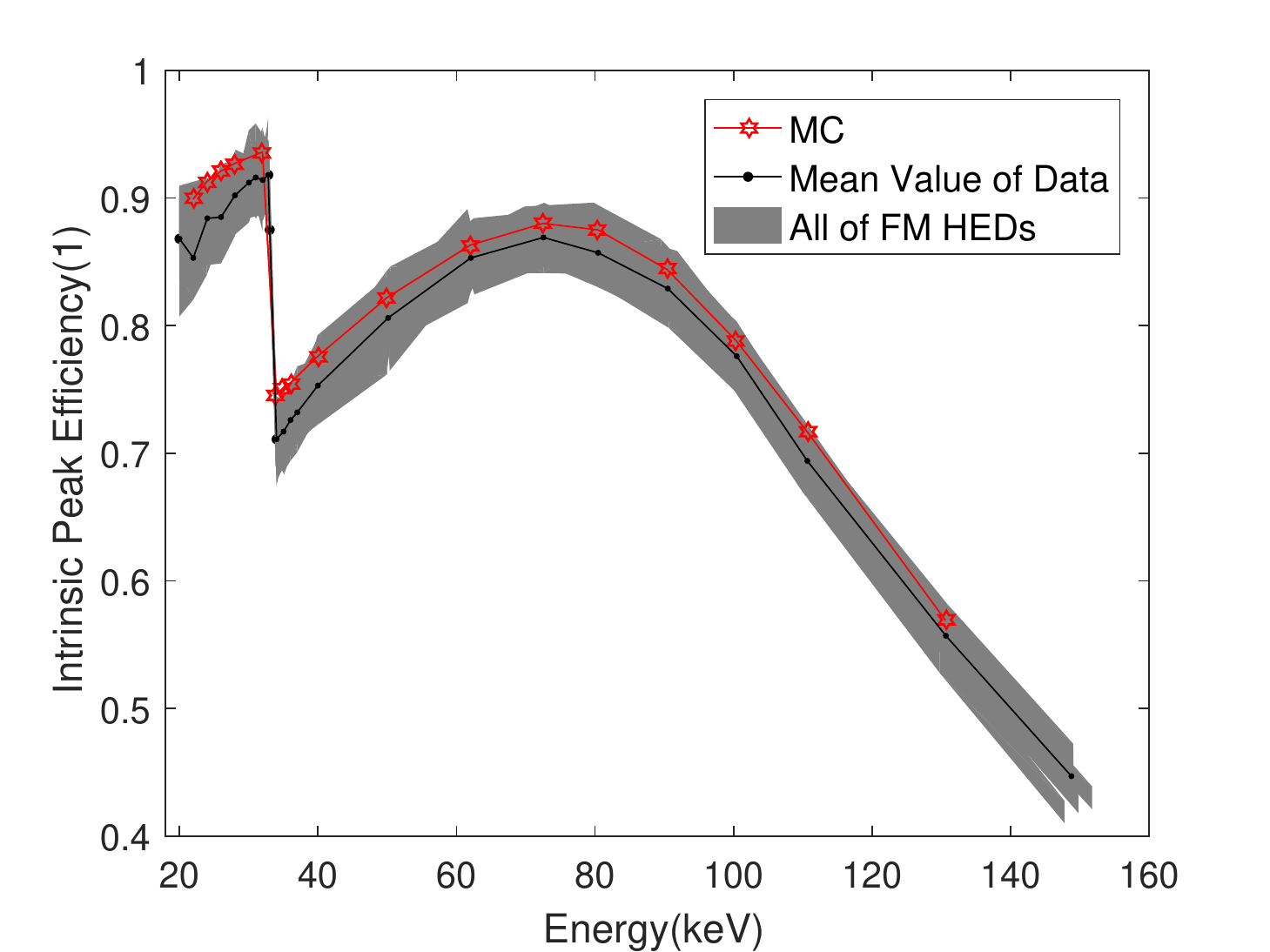}}
    \caption{HED Intrinsic peak efficiency as a function of photon energy. Top panel is for FM Z01-25 and the uncertainty are shown. Bottom panel is for all of FM HED/NaI , the average efficiency\,(black curve) and MC result\,(red curve) are also pointed}
    \label{fig:fig14 QDE}
\end{figure}

Where (1) $n(E_X)$ is the fitted full-energy peak area count rate of HED/NaI spectrum; (2) the beam flux $I(E_X)=\frac{n_{LEGe}(E_X)}{\varepsilon_{LEGe}(E_X)}$, where $ n_{LEGe}(E_X)$  is the count rate of the full-energy peak detected by the LEGe detector, $\varepsilon_{LEGe}(E_X)$  is the peak efficiency of LEGe at energy $E_X$ ; (3) $\kappa_I$  is the beam stability coefficient and set to 1 in our work.
Results for the intrinsic peak efficiency as a function of the energy for the HED/NaI are shown in Figure \ref{fig:fig14 QDE}. At system level, the effect of the anticoincidence shield system on the efficiency has also been tested with FM Z01-18 and one spare HED Z01-19. Then the effective area of the entire instrument as a function of energy will be simulated.

\subsection{Spatial Uniformity of HED} \label{sec:4.6 uniformity}
As already mentioned in Section \ref{sec:3.1 cal at HXCF}, the above performances of HED were determined at all energies by analyzing an integral spectrum taken in 27 positions of the detector's surface. The detector's spatial homogeneity was investigated by a more detailed scan of the surface area at 50\,keV. There 768 runs for FM Z01-1 and Z01-2, and 192 runs for the others. The AGC detector did not work in this test.

\begin{figure*}[!htbp]
  \centering
  \subfigure{
  \includegraphics[width=0.3\textwidth]{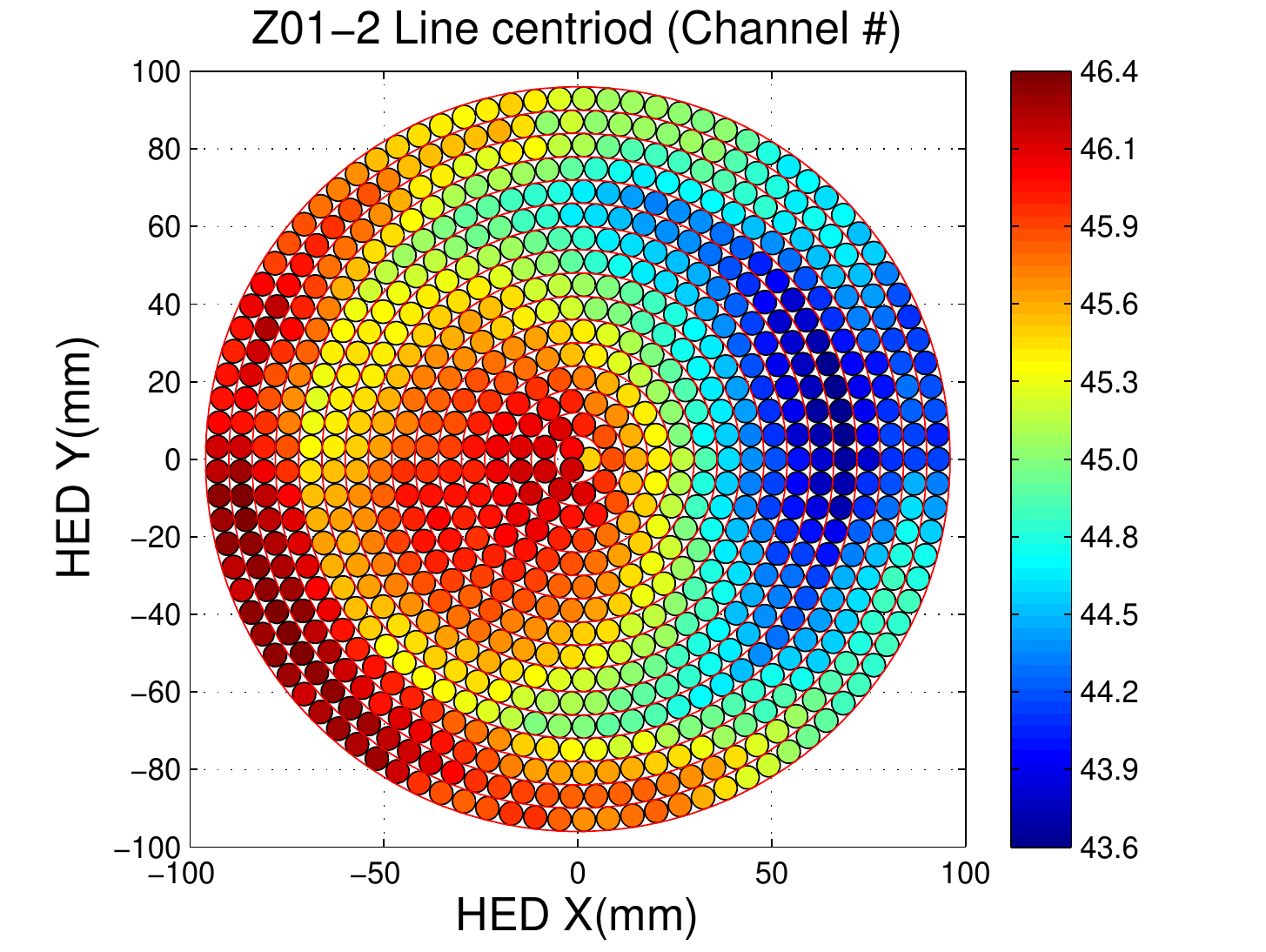}
  \includegraphics[width=0.3\textwidth]{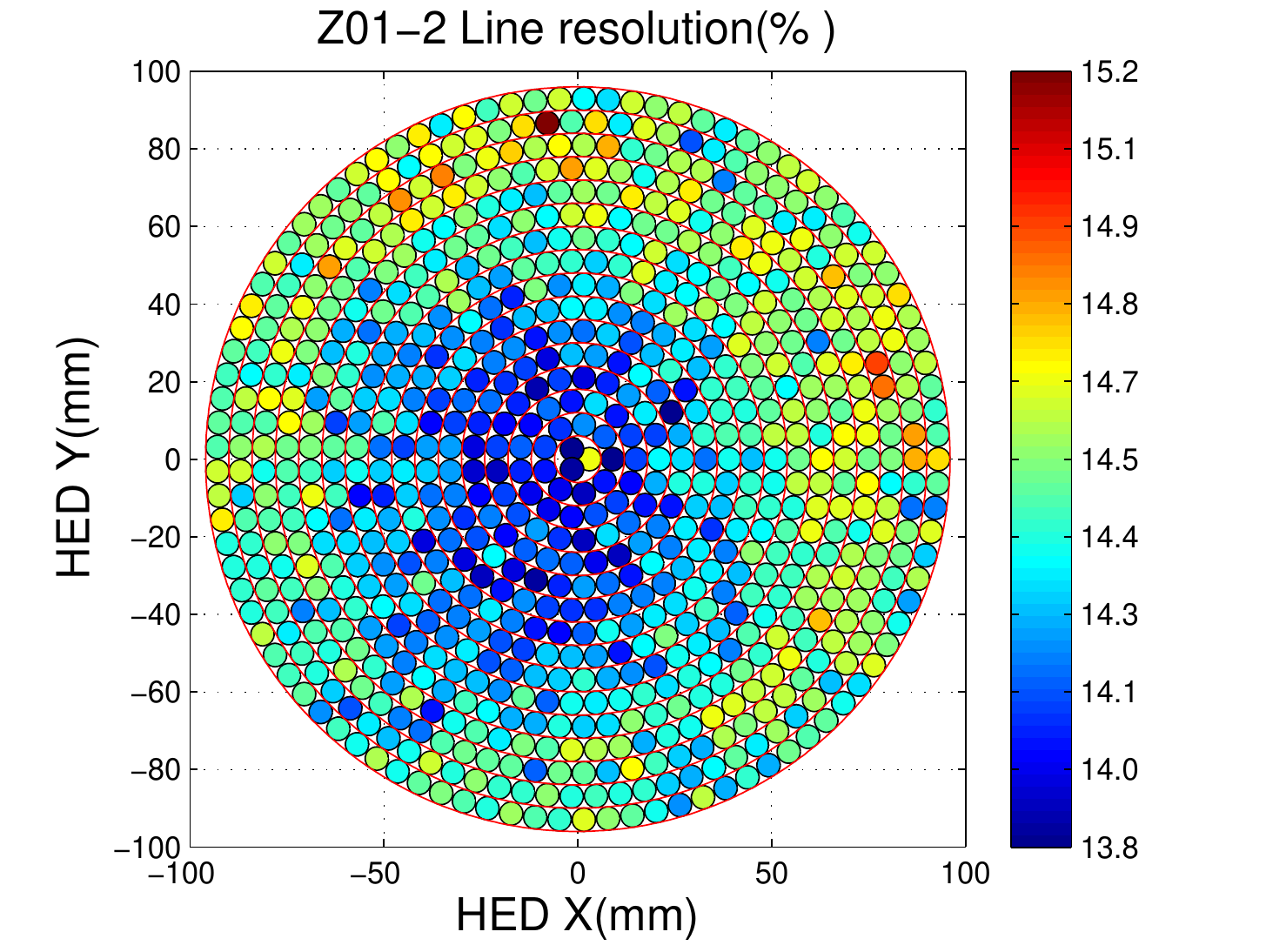}
  \includegraphics[width=0.3\textwidth]{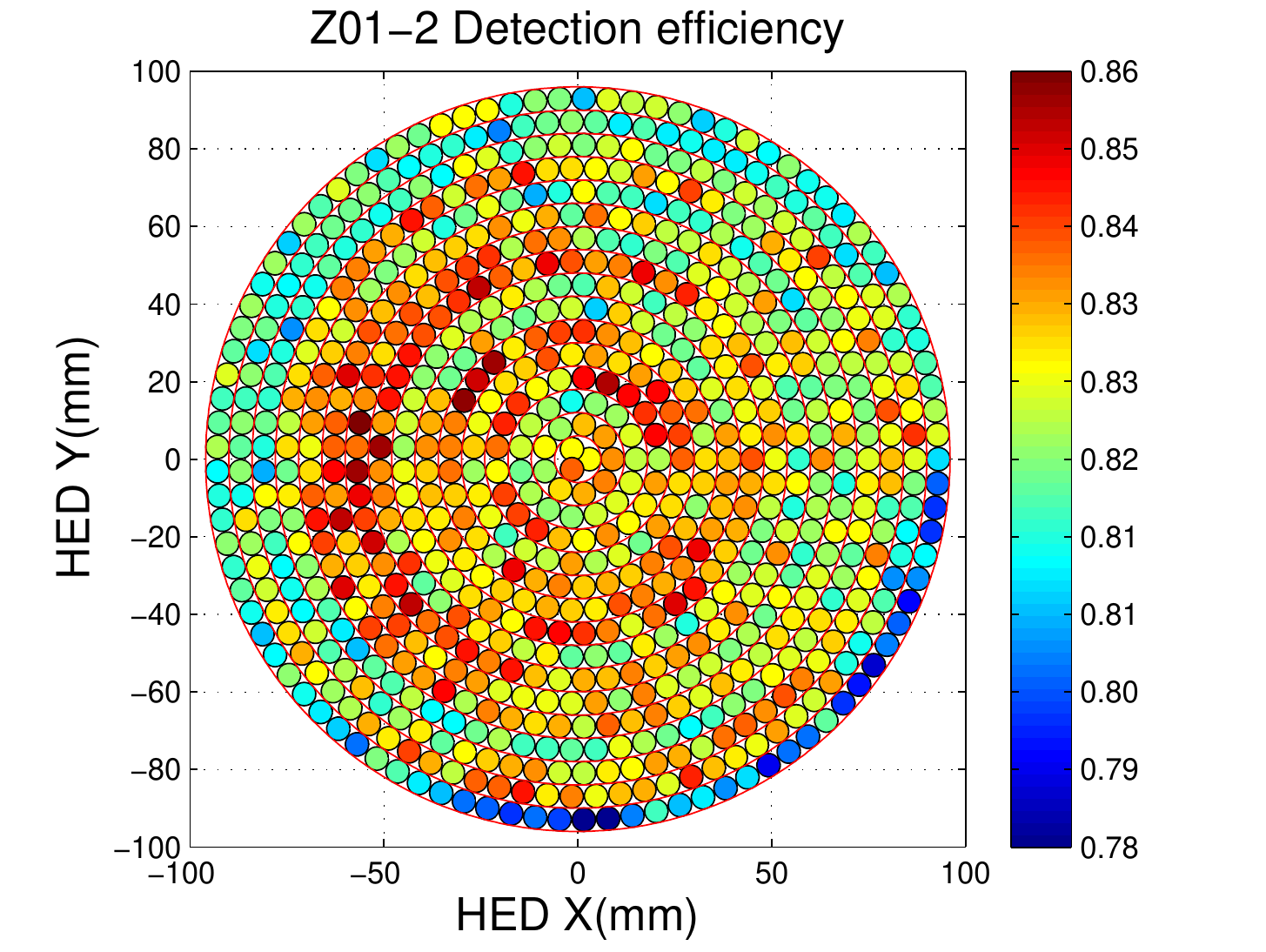}}
  \subfigure{
  \includegraphics[width=0.3\textwidth]{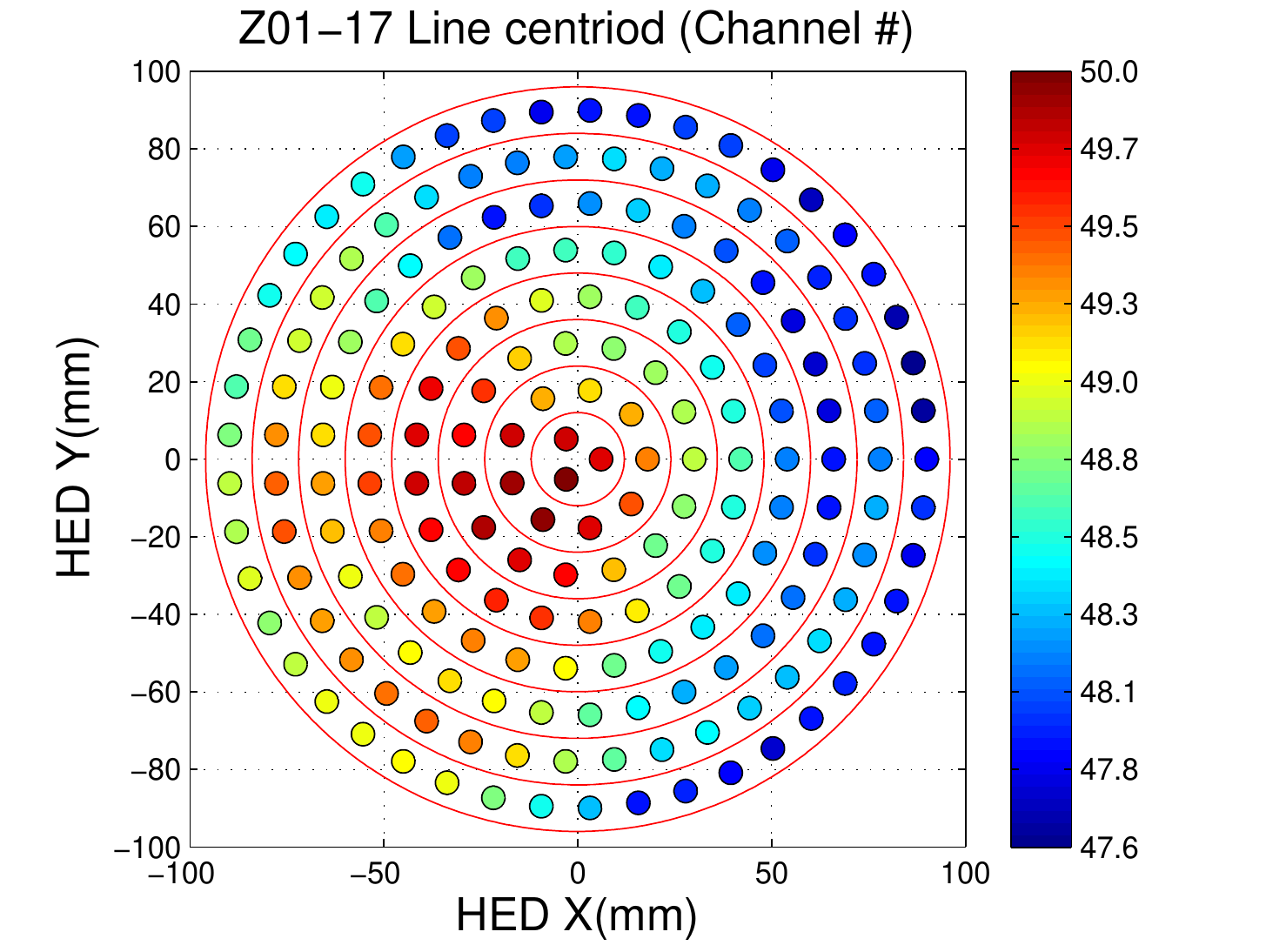}
  \includegraphics[width=0.3\textwidth]{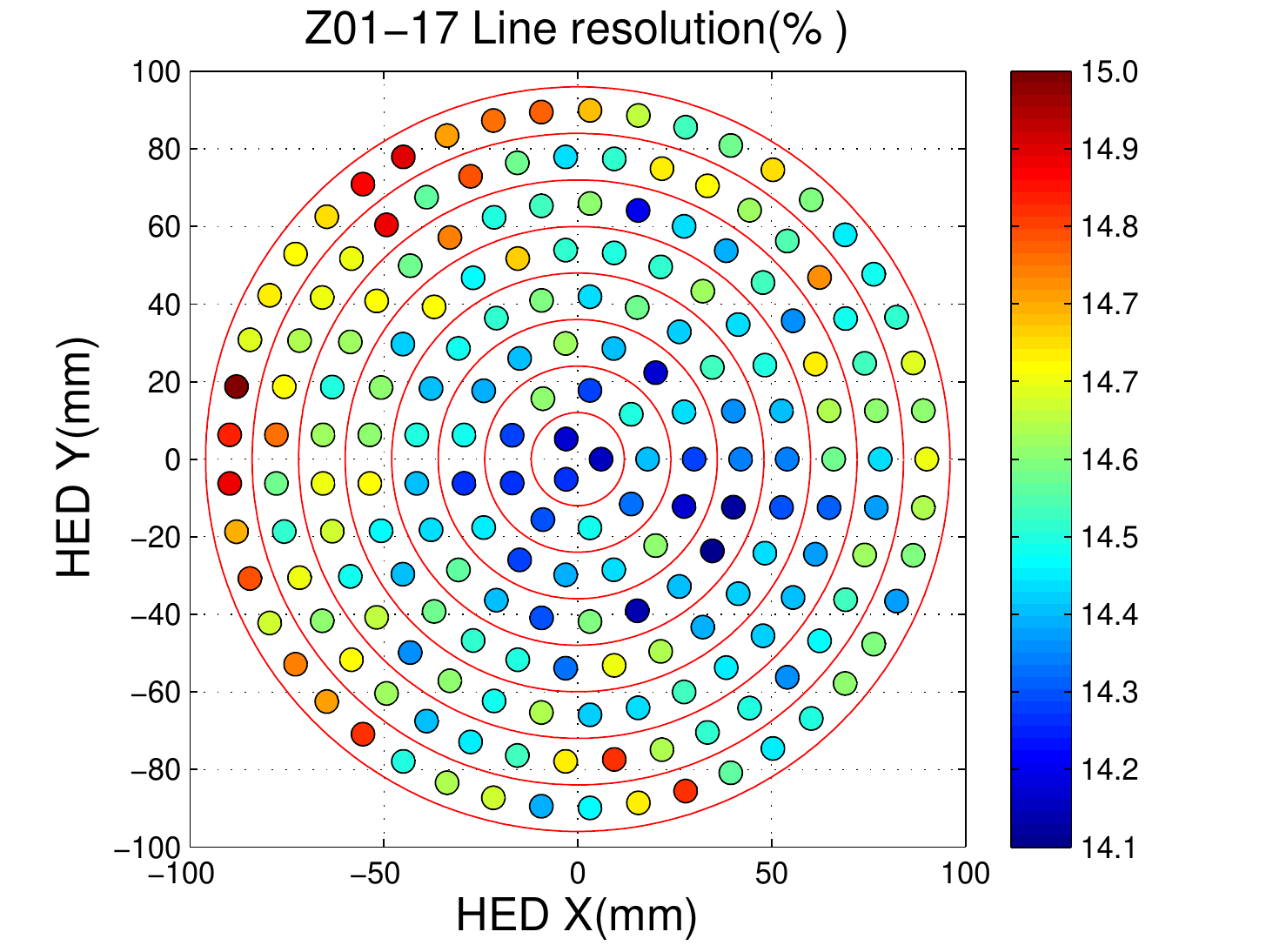}
  \includegraphics[width=0.3\textwidth]{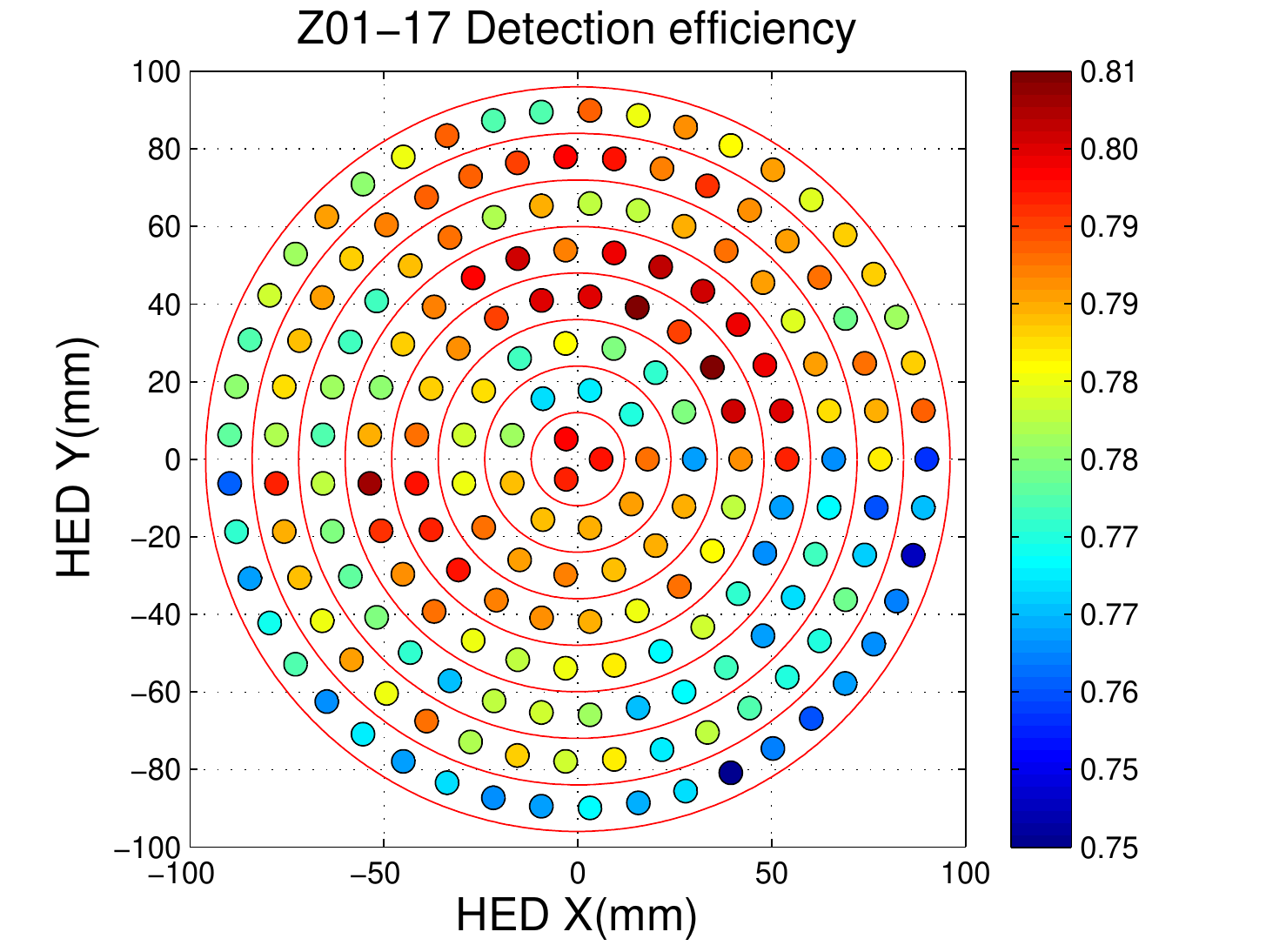}}
    \caption{Map of spatial uniformity results at 50\,keV for the HED Z01-2 and Z01-17(top and bottom panel, respectively). The performances of the full-energy peak as a function of the beam position are plotted: (1) the line-centroid (top panels); (2)the energy resolution (middle panels);(3) the peak efficiency (bottom panels).}
    \label{fig:fig15 Uniformity}
\end{figure*}

For each run the full-energy peak was analysed as previously described in section \ref{sec:4.2 peak fit}. The dependence of line-centroid (in channel), line resolution (in \%) and peak efficiency on the scan position in mm (for FM Z01-2 and Z01-17 ) is shown in Figure \ref{fig:fig15 Uniformity}. There are differences among detectors in the spatial uniformity, and two parameters are used to evaluate this performance: (1) $\frac{MAX-MIN}{MAX+MIN}$  , where the $MAX$  and $MIN$  is the maximum and minimum of the test results at different positions, respectively; (2) relative standard deviation(RSD). The former parameter of the line-centroid for all FM HEDs is between 0.027 and 0.044, and the latter is between 1.02\% and 1.75\%. The $\frac{MAX-MIN}{MAX+MIN}$  of the energy resolution for all FM detectors is 0.028$\sim$0.052 except for Z01-25 (is 0.104, because there is a small piece of area show poor resolution), and the RSD is 1.04\%$\sim$1.97\% but for Z01-25(is 2.38\%). The $\frac{MAX-MIN}{MAX+MIN}$ of the peak efficiency for all FM HEDs is 0.028$\sim$0.059, and the RSD is 1.13\%$\sim$2.2\%.

\section{Conclusions}
For simulating the physical detector response of the HE instrument of \emph{Insight}-HXMT, some on-ground calibration measurements were carried out with X-ray beam and radioactive sources. The channel-energy relations, energy resolutions and detection efficiencies of the single HED were characterized, these measurements results show a pretty good performance of HED, which meet the design requirements:

a) based on the fitted E-C relations, the energy range of each FM HED/NaI is 20 to $>\,300$\,keV, covers the required energy band (20 to 250\,keV);

b) energy resolution at 60keV of the HEDs is better than 15\%;

c) detection efficiencies of the HEDs follow the physical expectations.

d) the $\frac{MAX-MIN}{MAX+MIN}$ of the line centroid spatial homogeneity measured at 50keV for all HEDs is better than the required 0.05.

All those measurements and results have directly or indirectly contributed to the in-flight HE response matrix function determination, and which will be discussed in another paper.

%

\section*{ }

   \textit{This work was supported by the National Key R\&D Program of China (2016YFA0400800, 2016YFF0200802) , the Strategic Priority Research Program of the Chinese Academy of Sciences (Grant No. XDB23040400) and the National Natural Science Foundation of China under grant U1838104, U1838201, U1838202.}





%
%
%
%
\end{document}